\documentclass[fleqn,usenatbib]{mnras}

\usepackage{newtxtext,newtxmath}

\usepackage[T1]{fontenc}
\usepackage{ae,aecompl}

\usepackage{graphicx}	
\usepackage{amsmath}	
\usepackage{amssymb}	
\usepackage{bm}		
\usepackage{color}
\usepackage{xcolor}
\usepackage{soul}
\def\aj{AJ}%
\def\apj{ApJ}%
\def\apjl{ApJ}%
\def\apjs{ApJS}%
\def\ao{Appl.~Opt.}%
\def\aap{A\&A}%
\def\mnras{MNRAS}%
\def\prd{Phys.~Rev.~D}%
\def\prl{Phys.~Rev.~Lett.}%
\def\pasp{PASP}%
\def\pasj{PASJ}%
\def\jcap{JCAP}%
\def\physrep{Phys.~Rep.}%
\newcommand{\tbf}{\textbf}

\newcommand{\bea}{\begin{eqnarray}}
\newcommand{\be}{\begin{equation}}
\newcommand{\ben}{\begin{enumerate}}
\newcommand{\bi}{\begin{itemize}}
\newcommand{\eea}{\end{eqnarray}}
\newcommand{\ee}{\end{equation}}
\newcommand{\ei}{\end{itemize}}
\newcommand{\een}{\end{enumerate}}
\newcommand{\nn}{\nonumber}

\newcommand{\matC}{\mathbf C}

\newcommand{\like}{L}

\newcommand{\om}{\Omega_\mr m}
\newcommand{\omb}{\Omega_\mr b}
\newcommand{\sig}{\sigma_8}
\newcommand{\ns}{n_s}
\newcommand{\w}{w_0}
\newcommand{\wa}{w_a}

\newcommand{\mr}{\mathrm}

\usepackage{xspace}
\newcommand{\CL}{\textsc{CosmoLike}\xspace}

\newcommand*{\vertbar}{\rule[-1ex]{0.5pt}{2.5ex}}

\newcommand{\U}{\mathbf U}
\newcommand{\V}{\mathbf V}
\newcommand{\D}{\bm D}
\newcommand{\M}{\bm M}
\newcommand{\B}{\bm B}

\renewcommand{\P}{\mathbf P}
\newcommand{\C}{\mathbf C}

\newcommand{\invL}{\mathbf L^{-1}}
\renewcommand{\L}{\mathbf L}
\newcommand{\pco}{\bm p_{\rm co}}
\newcommand{\pnu}{\bm p_{\rm nu}}

\definecolor{azure}{rgb}{0.0, 0.5, 1.0}
\definecolor{darkgreen}{cmyk}{0.85,0.2,1.00,0.2}

\title[Cosmology with WFIRST - Synergies with LSST]{Cosmology with the Wide-Field Infrared Survey Telescope - Synergies with the Rubin Observatory Legacy Survey of Space and Time}

\author[Tim Eifler et al.]{Tim Eifler$^{1}$\thanks{E-mail:timeifler@arizona.edu}, 
Melanie Simet$^{2,3}$,
Elisabeth Krause$^{1,4}$,
Christopher Hirata$^{5}$,
Hung-Jin 
\newauthor
Huang$^{1}$,
Xiao Fang$^{1}$,
Vivian Miranda$^{1}$,
Rachel Mandelbaum$^{6}$, 
Cyrille Doux$^{7}$,
Chen 
\newauthor
Heinrich$^{3}$,
Eric Huff$^{3}$,
Hironao Miyatake$^{3,8,9,10}$,
Shoubaneh Hemmati$^{3}$, 
Jiachuan Xu$^{1}$,
\newauthor 
Paul Rogozenski$^{4}$, 
Peter Capak$^{12}$,
Ami Choi$^{5}$,
Olivier Dor\'e$^{3,11}$,
Bhuvnesh Jain$^{7}$,
\newauthor
Mike Jarvis$^{7}$,
Niall MacCrann$^{5}$,
Dan Masters$^{3}$, 
Eduardo Rozo$^{4}$, 
David N. Spergel$^{13,14}$,
\newauthor
Michael Troxel$^{15}$,
Anja von der Linden$^{16}$,
Yun Wang$^{12}$,
David H. Weinberg$^{5}$,
\newauthor
Lukas Wenzl$^{17}$,
Hao-Yi Wu$^{5}$
\\
$^{1}$ Department of Astronomy/Steward Observatory, University of Arizona, 933 North Cherry Avenue, Tucson, AZ 85721-0065, USA \\
$^{2}$ University of California Riverside, 900 University Ave, Riverside, CA 92521, USA \\
$^{3}$ Jet Propulsion Laboratory, California Institute of Technology, Pasadena, CA 91109, USA\\
$^{4}$ Department of Physics, University of Arizona, , 1118 E Fourth Str, AZ 85721, USA \\
$^{5}$Center for Cosmology and AstroParticle Physics, The Ohio State University, 191 West Woodruff Avenue, Columbus, Ohio 43210, USA\\
$^{6}$ McWilliams Center for Cosmology, Department of Physics, Carnegie Mellon University, Pittsburgh, PA 15213, USA \\
$^{7}$Department of Physics and Astronomy, University of Pennsylvania, Philadelphia, PA 19104, USA \\
$^{8}$Institute for Advanced Research, Nagoya University, Nagoya 464-8601, Japan\\
$^{9}$Division of Physics and Astrophysical Science, Graduate School of Science, Nagoya University, Nagoya 464-8602, Japan \\
$^{10}$ Kavli IPMU~(WPI), UTIAS, The University of Tokyo, Chiba 277- 8583, Japan\\
$^{11}$California Institute of Technology, 1200 E. California Blvd., Pasadena, CA 91125, USA\\
$^{12}$IPAC, California Institute of Technology, Pasadena, CA 91125, USA\\
$^{13}$Center for Computational Astrophysics, Flatiron Institute, NY NY 10010, USA\\
$^{14}$Department of Astrophysical Sciences, Princeton University, Princeton NJ 08544, USA\\
$^{15}$Department of Physics, Duke University, Durham, NC 27708, USA\\
$^{16}$Department of Physics and Astronomy, Stony Brook University, Stony Brook, NY 11794, USA\\
$^{17}$Department of Astronomy, Cornell University, Ithaca, NY 14853, USA
}

\date{Accepted XXX. Received YYY; in original form ZZZ}

\pubyear{2015}

\begin{document}
\label{firstpage}
\pagerange{\pageref{firstpage}--\pageref{lastpage}}
\maketitle

\begin{abstract} 
We explore synergies between the space-based Wide-Field Infrared Survey Telescope (WFIRST) and the ground-based Rubin Observatory Legacy Survey of Space and Time (LSST). In particular, we consider a scenario where the currently envisioned survey strategy for WFIRST's High Latitude Survey (HLS), i.e., 2000 deg$^2$ in four narrow photometric bands is altered in favor of a strategy that combines rapid coverage of the LSST area (to full LSST depth) in one band. 
We find that a 5-month WFIRST survey in the W-band can cover the full LSST survey area providing high-resolution imaging for $>$95\% of the LSST Year 10 gold galaxy sample. We explore a second, more ambitious scenario where WFIRST spends 1.5 years covering the LSST area. For this second scenario we quantify the constraining power on dark energy equation of state parameters from a joint weak lensing and galaxy clustering analysis, and compare it to an LSST-only survey and to the Reference WFIRST HLS survey. 
Our survey simulations are based on the WFIRST exposure time calculator and redshift distributions from the CANDELS catalog. Our statistical uncertainties account for higher-order correlations of the density field, and we include a wide range of systematic effects, such as uncertainties in shape and redshift measurements, and modeling uncertainties of astrophysical systematics, such as galaxy bias, intrinsic galaxy alignment, and baryonic physics. Assuming the 5-month WFIRST wide scenario, we find a significant increase in constraining power for the joint LSST+WFIRST wide survey compared to LSST Y10 (FoM$_\mr{Wwide}$= 2.4 FoM$_\mr{LSST}$) and compared to LSST+WFIRST HLS (FoM$_\mr{Wwide}$= 5.5 FoM$_\mr{HLS}$).
\end{abstract}




\section{Introduction}
\label{sec:intro}
Observing the large-scale structure in our Universe provides a wealth of cosmological information and allows us to constrain fundamental physics questions such as the nature of dark energy, the mass and number of species of neutrinos, possible modifications to general relativity as a function of scale or environment, or the nature of dark matter interactions. 

Early results from the Dark Energy Survey \citep[DES\footnote{www.darkenergysurvey.org/},][]{kez17, tmz18, DES3x2,DES5x2}, the Kilo-Degree Survey \citep[KiDS\footnote{http://www.astro-wise.org/projects/KIDS/},][]{ujs18,hkb18,khd19}, and the Hyper Suprime Cam Subaru Strategic Program \citep[HSC\footnote{http://www.naoj.org/Projects/HSC/HSCProject.html},][]{hoh19,hsm19} have demonstrated the feasibility of complex (multi-probe) analyses from photometric data. Within the $\Lambda$CDM model, these surveys show a $\sim$2-$\sigma$ tension/agreement with Cosmic Microwave Background (CMB) measurements from the Planck satellite \citep{Planckcosmo18}, which sets the stage for exciting near future discoveries when the full data of DES, HSC, KiDS is analyzed.  

This discovery potential will increase significantly in the coming decade when photometric data from Stage 4 surveys become available \citep[see][for a review]{wme13}. The Rubin Observatory Legacy Survey of Space and Time \citep[LSST\footnote{https://www.lsst.org/},][]{LSST19}, Euclid\footnote{https://sci.esa.int/web/euclid} \citep{laa11}, the Spectro-Photometer for the History of the Universe, Epoch of Reionization, and Ices Explorer \citep[SPHEREx\footnote{http://spherex.caltech.edu/},][]{dba14}, and the Wide Field Infrared Survey Telescope \citep[WFIRST\footnote{https://wfirst.gsfc.nasa.gov/},][]{sgb15} complement each other in terms of area, depth, wavelength, and resolution and will provide a wealth of cosmological data to be mined by the community. The concert of cosmological endeavors in the 2020s also includes several spectroscopic experiments, e.g. the Dark Energy Spectroscopic Instrument \citep[DESI, ][]{DESI16}, the Prime Focus Spectrograph \citep[PFS, ][]{tec14}, the 4-metre Multi-Object Spectroscopic Telescope \citep[4MOST, ][]{4MOST19}, and of course the spectroscopic components of Euclid and WFIRST. Together with the next generation of CMB surveys, such as the Simons Observatory \citep[SO, ][]{SO19} and CMB-S4 \citep{CMBS4} these experiments form a highly synergistic ensemble of data sets to study the physics of our Universe.

In this paper we focus on synergies between WFIRST and LSST, specifically we explore alterations to the WFIRST observing strategy that maximize said synergies for a joint (photometric) clustering and weak lensing (so-called 3x2pt) analysis. A joint LSST+WFIRST data set will enable a variety of other cosmological probe combinations and we refer the 
reader to a companion paper \citep{emk20} for analysis strategies that include information from galaxy clusters, spectroscopic clustering, and Type Ia supernova.

\paragraph*{Vera C. Rubin Observatory} will start commissioning and science verification in 2021 and is scheduled to be fully operational in 2022. The LSST is best described as a wide-fast-deep-repeat survey that will collect $\sim$15 TB of imaging data per day. This daily delivery of high quality data to the science community will continue for at least 10 years, i.e. until 2032.  Thanks to its 9.6 deg$^2$ field-of-view, 3.2 Gigapixel camera LSST can cover 10,000 deg$^2$ in one of its 6 bands (320nm-1050nm) every 3 nights. The short exposure time of 2$\times$15 s and rapid mapping of the night sky is ideal to find transients and solar objects; for cosmologists, LSST offers a wide survey of 18,000 deg$^2$, single exposure depth of 24.7 r-band magnitude (5$\sigma$ point source), and a design optimized for photometric homogeneity and astrometric accuracy. The final data set will encompass >20 billion galaxies and a coadded map down to a depth of r-band magnitude 27.5. 

\paragraph*{The Wide Field Infrared Survey Telescope} \citep{sgb15} is scheduled to launch mid 2020s and it is probably best described as a multi-purpose space observatory, with a 100 times larger field-of-view than the Hubble Space Telescope. WFIRST's versatile capabilities address science cases ranging from exoplanets to galaxy evolution to fundamental physics. WFIRST's High Latitude Survey \citep[HLS,][]{WFIRSTrep} is designed to constrain dark energy evolution and deviations from GR with excellent control of systematics via space-quality imaging, photometry across 4 near-infrared (NIR) bands, a 0.28 deg$^2$ field of view with a 0.11 arcsec pixel scale, and $\sim$600 resolution grism spectroscopy. The Reference\footnote{The Reference survey is being used to assess whether the WFIRST design meets requirements. No decisions have been made regarding what survey will actually be executed.} design of the HLS assumes a duration of 1.6 years covering 2,000 deg$^2$ with the imaging and spectroscopic modes. This strategy enables deep spectroscopic galaxy redshift measurements over the same survey area as the imaging survey.

\renewcommand{\arraystretch}{1.3}
\begin{table*}
\caption{Technical specifications characterizing the photometric data sets from LSST and WFIRST. Sensitivity refers to 5$\sigma$ point source depth (please see table for exact exposure times that correspond to the depth numbers). We note that the exact strategy for LSST and WFIRST exposure duration is still being discussed.} 
\begin{center}
\begin{tabular*}{0.99\textwidth}{@{\extracolsep{\fill}}| c | c c c c c c c |}
\hline
\hline
\multicolumn{8}{|c|}{\tbf{WFIRST parameters, following \citet{sgb15}}} \\
\hline 
\multicolumn{2}{|c }{\tbf{Telescope Aperture}}  & \multicolumn{2}{|c }{\tbf{Field of View}} & \multicolumn{2}{|c }{\tbf{Pixel Scale}} & \multicolumn{2}{|c|}{\tbf{Wavelength Range}} \\
\multicolumn{2}{|c }{\tbf{2.4 meter}}  & \multicolumn{2}{|c }{\tbf{0.28 sq deg}} & \multicolumn{2}{|c }{\tbf{0.11 arcsec}} & \multicolumn{2}{|c|}{\tbf{0.5-2.0 $\mu$m}} \\
\hline 
\hline 
Filters & R062 & Z087& Y106 & J129 & H158 & F184 & W146 \\
Wavelength ($\mu$m) & 0.48 - 0.76 & 0.76 - 0.98 & 0.93 - 1.19 & 1.13 - 1.45 & 1.38 - 1.77 & 1.68 - 2.00 & 0.93 -2.00 \\
Sensitivity (AB mag \tbf{in 1h}) &28.5 & 28.02 & 27.95 & 27.87 & 27.81 & 27.32 & 28.33 \\
\hline 
\hline
\end{tabular*}

\vspace{0.3cm}

\begin{tabular*}{0.99 \textwidth}{@{\extracolsep{\fill}}| c | c c c c c c |}
\hline
\hline
\multicolumn{7}{|c|}{\tbf{LSST parameters (\citet{DESC_SRD18} and \citet{LSST19}).}} \\
\hline 
\multicolumn{2}{|c }{\tbf{Telescope Aperture}}  & \multicolumn{2}{|c }{\tbf{Field of View}} & \multicolumn{2}{|c|}{\tbf{Pixel Scale}} & \tbf{Wavelength Range} \\
\multicolumn{2}{|c }{\tbf{6.67 meter (effective)}}  & \multicolumn{2}{|c }{\tbf{9.62 sq deg}} & \multicolumn{2}{|c| }{\tbf{0.2 arcsec}} & \tbf{0.32 - 1.0 $\mu$m} \\
\hline 
\hline 
Filters & u & g & r & i & z & y \\
Wavelength ($\mu$m) & 0.32 - 0.40 & 0.40 - 0.55 & 0.55 - 0.69 & 0.69 - 0.82 & 0.82 - 0.92 & 0.92 - 1.0 \\
Sensitivity (AB mag, Full Survey) &25.30&26.84& 27.04& 26.35& 25.22& 24.47\\
\hline 
\hline
\end{tabular*}
\end{center}
\label{tab:specs}
\end{table*}
\renewcommand{\arraystretch}{1.0}

Table \ref{tab:specs} summarizes some of the main technical specifications of WFIRST and LSST. Most obvious is the complementarity in multi-wavelength information with LSST pushing towards the blue end in the visible whereas WFIRST extends the color information into the near-infrared. We note that while the current HLS survey design plans on using the YJHF bands, there are several other filters available, in particular the W-band ranging from 0.93-2.0 microns, which was introduced for the microlensing survey \citep{pgk19}, and also bands that overlap with LSST towards the red part of the visible spectrum >0.5 microns. 

In this paper we comment on several scenarios where a single WFIRST band is used for a rapid, deep coverage of the LSST area. The combination of color information down to 320 nm from the ground, and high-resolution infrared coverage from space over 18,000 deg$^2$ can potentially unlock a new level of precision for the core science cases of both experiments, if and only if, systematics control is maintained at the new level. Throughout the paper we refer to this single band concept as \textit{WFIRST wide} and we will contrast it with the Reference 4 NIR-band 2,000 deg$^2$ survey design, which we refer to as \textit{WFIRST HLS}. We iterate that both concepts, WFIRST wide and WFIRST HLS, include LSST data as part of their analysis and should technically be labeled LSST+WFIRST.

\section{Multi-probe analysis basics - Combining Weak Lensing and Galaxy Clustering}
\label{sec:basics}
\subsection{Forecasting Philosophy}

A large variety of cosmological probes can be extracted from wide-field imaging and spectroscopic WFIRST observations, such as various forms of weak lensing (e.g., cosmic shear using second and higher order summary statistics, magnification, galaxy-galaxy lensing, cluster weak lensing, and shear peak statistics), galaxy clustering (spectroscopic 3D and projected 2D), galaxy cluster number counts, CMB-lensing cross-correlations, Baryon Acoustic Oscillations, and redshift space distortions. 

Statements about the constraining power of different survey concepts are a function of the exact ingredients for the data vector, which does not just include the combination of probes but also the exact scales and redshifts that are included, the systematics considered, the parameterization and priors of said systematics, and the priors and parameters of the cosmological model of interest. 

Our forecasting philosophy is driven by the intent to model a realistic WFIRST and LSST 3x2pt analysis (joint weak lensing and galaxy clustering). A realistic assessment of the constraining power means to compute realistic error bars, which requires accurately computing three main components of a likelihood analysis: 1) statistical uncertainties, 2) systematic uncertainties, 3) inference aspects of the likelihood analysis. 

Regarding the statistical uncertainties, we use analytic, non-Gaussian covariance matrices that include higher order moments of the density field and super-sample variance and that have been compared to numerical simulations (albeit only in less constraining cases). We also use realistic survey simulations (see Sect. \ref{sec:surveysims}) to quantify the number density and redshift distributions of source and lens sample and to compute a realistic survey area for a given scenario.   
Regarding systematic uncertainties, we include observational uncertainties such as shear calibration uncertainties and photo-z errors for lens and source sample. We parameterize a variety of astrophysics uncertainties: galaxy bias uncertainties are captured through a combination of scale cuts and nuisance parameters; similarly we include uncertainties from intrinsic galaxy alignment and baryonic physics modeling again through a combination of nuisance parameters and scale cuts. The exact implementation is described in Sect. \ref{sec:like}.

Regarding robust inference, we opted for simulated MCMC analyses instead of Fisher matrices in order to avoid numerical uncertainties in computing high-dimensional (56 dimensions spanning cosmological and nuisance parameters) derivatives of the observables and in order to capture posterior probabilities beyond the multivariate Gaussian.

We employ the \textsc{CosmoLike} framework \citep{eks14, kre17} for the simulated likelihood analyses, however we do not use its most sophisticated modules to model the density power spectrum and the subsequent projected power spectra. For example, we do not use a Boltzmann code to compute the initial power spectrum nor do we include reduced shear, non-Limber, or magnification in the calculation of the summary statistics. We reason that since the (noise-free) data vector is computed with the exact same \textsc{CosmoLike} module, the per cent uncertainty arising from these approximations do not alter our conclusions regarding the relative merits of different survey options. Generally speaking, we argue that modeling uncertainties can be ignored for the purpose of quantifying constraining power unless they contribute to the error budget at a level comparable to our main error sources (statistical uncertainties, shear calibration, photo-z uncertainties, galaxy bias, intrinsic alignment, baryonic physics).

Adopting this strategy has the major advantage that we can use the much faster forecasting routines of \CL, which speeds up our simulated likelihood analysis by a factor of $\sim$20, from 10s per MCMC step to $\sim$0.5s . This allows us to compute long, converged chains, and it enables a future in-depth survey scenario study that varies the baseline choices and priors presented in Table \ref{tab:params}.  

\subsection{Modeling Details}
In this section we summarize the computation of angular (cross) power spectra for the different probes closely following the notation in \citet{kre17}; a more detailed derivation can be found in \citet{huj04}. We use capital Roman subscripts to denote observables, $A,B\in \left\{ \kappa,\delta_{\mathrm{g}}\right\}$, where $\kappa$ refers to the lensing of source galaxies and $\delta_{\mathrm{g}}$ is the density contrast of the lens galaxy sample. 

We note the frequent confusion originating from the terms ``lens sample'' and ``lensed galaxy sample'' and clarify that we only use the position information of the lens sample. Specifically, the auto-correlation of positions of lens sample galaxies will later form the galaxy clustering part of the 3x2pt data vector, whereas the autocorrelation of lensed galaxy shapes, that is the cosmic shear part of the data vector, is computed using the source galaxy sample.

The angular power spectrum between redshift bin $i$ of observable $A$ and redshift bin $j$ of observables $B$ at projected Fourier mode $l$, $C_{AB}^{ij}(l)$ is computed using the Limber and flat sky approximations:
\be
\label{eq:projected}
C_{AB}^{ij}(l) = \int d\chi \frac{q_A^i(\chi)q_B^j(\chi)}{\chi^2}P_{AB}(k=l/\chi,\;z(\chi)),
\ee
where $\chi$ is the comoving distance, $q_A^i(\chi)$ are weight functions of the different observables given in Eqs.~(\ref{eq:qg}-\ref{eq:qkappa}), and $P_{AB}(k,z)$ the three dimensional, probe-specific power spectra detailed below. We note that using the Limber approximation can have significant impact on the parameter estimation \citep[e.g., see][]{2019arXiv191111947F} when analyzing actual data, however it is not a concern for our simulated analyses.

The weight function for the projected galaxy density in redshift bin $i$, $q_{\delta_{\mathrm{g}}}^i(\chi)$, is given by the normalized comoving distance probability of galaxies in this redshift bin
\be
\label{eq:qg}
q_{\delta_{\mathrm{g}}}^i(\chi) =\frac{n_{\mathrm{lens}}^i(z(\chi)) }{\bar{n}_{\mathrm{lens}}^i}\frac{dz}{d\chi}\,,
\ee
with $n_{\mathrm{lens}}^i(z)$ the redshift distribution of galaxies in (photometric) galaxy redshift bin $i$ (cf. Eq. (\ref{eq:photoz})), and $\bar{n}_{\mathrm{lens}}^i$ the areal number densities of lens galaxies in this redshift bin. 

For the convergence field, the weight function $q_\kappa^{i}(\chi)$ is the lens efficiency, 
\be
\label{eq:qkappa}
q_\kappa^{i}(\chi) = \frac{3 H_0^2 \Omega_m }{2 \mathrm{c}^2}\frac{\chi}{a(\chi)}\int_\chi^{\chi_{\mr h}} \mr d \chi' \frac{n_{\mathrm{source}}^{i} (z(\chi')) dz/d\chi'}{\bar{n}_{\mathrm{source}}^{i}} \frac{\chi'-\chi}{\chi'} \,,
\ee
where $n_{\mathrm{source}}^{i} (z)$ denotes the redshift distribution of source galaxies in (photometric) source redshift bin $i$ (cf. Eq. (\ref{eq:photoz})), $\bar{n}_{\mathrm{source}}^i$ the areal angular  number densities of source galaxies in this redshift bin $i$, and $a(\chi)$ is the scale factor.

We model the probe specific three-dimensional power spectra $P_{AB}(k,z)$ as a function of the nonlinear matter density power spectrum $P_{\delta\delta}(k,z)$, for which we use the \citet{tsn12} fitting formula
\bea
P_{\kappa \kappa}(k,z) = P_{\delta\delta}(k,z) \,, \\
P_{\delta_g \kappa}(k,z) = b_g(z) P_{\delta\delta}(k,z) \,, \\
P_{\delta_g \delta_g}(k,z) = b^2_g(z) P_{\delta\delta}(k,z) \,.
\eea
We only consider the large-scale galaxy distribution and assume that the galaxy density contrast on these scales can be approximated as the nonlinear matter density contrast times an effective galaxy bias parameter $b_g(z)$.
We defer the exact implementation to our systematics discussion in Sect. \ref{sec:like}.

\subsection{Covariance and Inference Details}
The multi-probe data vector, denoted as $\D$, is computed at the fiducial cosmology and systematics parameter values (see Table \ref{tab:params}). The same parameters are assumed in the computation of the non-Gaussian covariance matrix, $\matC$. Given that this covariance matrix is calculated analytically, it is not an estimated quantity derived from either simulations or measured data. As a quantity that is free of estimator noise analytical covariance matrices can be inverted directly and do not require large amounts of realizations for the inverse to be precise \citep[see e.g.,][for details on the number of realizations and alternative ideas]{tjk13,dos13,fre18}. 

We sample the joint parameter space of cosmological $\pco$ and nuisance parameters $\pnu$, the latter describing our systematic uncertainties, and parameterize the joint likelihood as a multivariate Gaussian 
\be
\label{eq:like}
\like (\D| \pco, \pnu) = N \, \times \, \exp \biggl( -\frac{1}{2} \underbrace{\left[ (\D -\M)^t \, \matC^{-1} \, (\D-\M) \right]}_{\chi^2(\pco, \pnu)}  \biggr) \,.
\ee
The model vector $\M$ is a function of cosmology and nuisance parameters, i.e. $\M=\M(\pco, \pnu)$ and the normalization constant $N=(2 \pi)^{-\frac{n}{2}} |C|^{-\frac{1}{2}}$ can be ignored under the assumption that the covariance is constant while the MCMC steps through the parameter space. 

We calculate the covariance of two angular power spectra as the sum of the Gaussian covariance, $ \mr{Cov^G}$, and non-Gaussian covariance in the absence of survey window effects $\mr{Cov^{NG}}$, and the super-sample covariance, $\mr{Cov^{SSC}}$, which describes the uncertainty induced by large-scale density modes outside the survey window

\bea
\mr{Cov}\left( C_{AB}^{ij} (l_1), C_{AB}^{ij}(l_2) \right)  &=& \mr{Cov^G}\left( C_{AB}^{ij}(l_1), C_{AB}^{ij}(l_2) \right) \nn \\
&+&\mr{Cov^{NG}}\left( C_{AB}^{ij}(l_1), C_{AB}^{ij}(l_2) \right)  \nn \\
&+&\mr{Cov^{SSC}}\left( C_{AB}^{ij}(l_1), C_{AB}^{ij}(l_2) \right)\,.
\eea
We refer the interested reader to the Appendix (Eqs: A2-A14) of \citet{kre17} for the exact implementation of the 3 covariance terms.

However, the assumption of a covariance matrix $\matC$ that remains constant throughout the MCMC process, with the input parameters perfectly known, is illogical given that it is exactly said parameters that we aim to constrain in a likelihood analysis. \cite{esh09} explore the idea of varying the covariance as a function of parameters, similar to the way one varies the model data vector. \cite{car13} shows that using a parameter dependent covariance in combination with a Gaussian likelihood function violates the Cram\'er-Rao bound and recommends not to use this combination in a likelihood analysis. This recommendation has led e.g., DES likelihood analyses to adapt a second recommendation of \cite{esh09}: an iterative likelihood analysis approach, where the parameters of the covariance are updated in a new likelihood analysis based on the best-fit parameters of the previous run.  

We follow this approach and assume a multivariate Gaussian likelihood as described in Eq. (\ref{eq:like}) with a constant covariance matrix computed at the fiducial parameters for all the surveys considered. 
    
Given the likelihood function we compute the posterior probability in parameter space from Bayes' theorem and we sample the parameter space using \texttt{emcee\footnote{\url{https://emcee.readthedocs.io/en/stable/}}} \citep{fhg13}, which is based on the  affine-invariant  sampler of  \cite{gow10}, and which can be parallelized with either MPI or shared memory multiprocessing.

\section{Simulating WFIRST and LSST surveys}
\label{sec:surveysims}
\subsection{LSST}
In order to incorporate the latest analysis choices for a 3x2pt LSST analysis we closely follow the LSST~DESC Science Requirements Document v1, published as \citet[][]{DESC_SRD18} (D18 from hereon), which reflects a conservative consensus choice across the LSST~DESC. In D18 survey depths are based on the \texttt{OpSim} software\footnote{\url{https://github.com/lsst/sims_operations}}, specifically the OpSim v3 \texttt{minion\_1016} run. 
The scenario described in D18 has median depth (5$\sigma$ point-source detection across the survey) of 25.30, 26.84, 27.04, 26.35, 25.22, 24.47 in ugrizy after 10 years.  
These numbers are determined after discarding areas with limiting i-band magnitude (AB) i-mag<26 
to homogenize the depth across the whole survey. LSST dark energy science will exclude data from regions near a Galactic latitude of zero and likely discard areas with E(B-V)>0.2 . In D18 the resulting area after this homogenization process is 14,300 deg$^2$. For the purpose of this paper we however assume the more ambitious survey area of 18,000 deg$^2$, which is more demanding in terms of WFIRST observing time, if WFIRST were to cover the full LSST footprint.

LSST is evaluating a variety of different survey strategies \citep[see e.g.,][]{OSTF18} with possible survey areas ranging from below 15,000 deg$^2$ to more than 25,000 deg$^2$. 
Our 18,000 deg$^2$ extragalactic sky coverage scenario is well within the range of options although we note that the survey depth choices of this paper can only be achieved if LSST would shift observing time from the low Galactic latitudes to the extragalactic survey.  

\begin{figure}
\includegraphics[width=8.1cm]{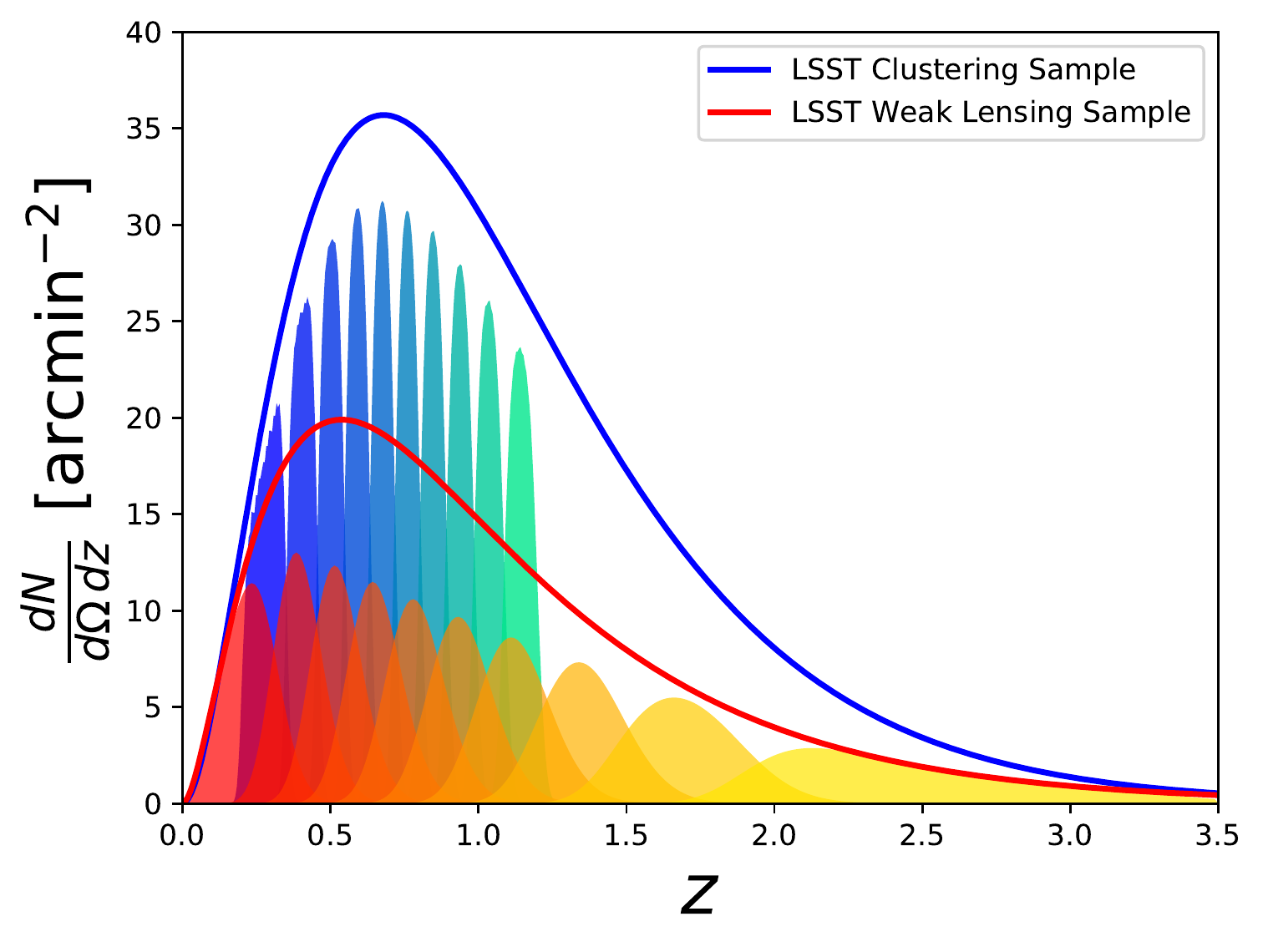}
\caption{LSST redshift distribution with 10 tomographic lens bins (blue) and 10 tomographic source bins (red), based on the ingredients from D18. We note the 10 lens bins are at lower redshift ($\le 1.2$) and are more narrow compared to the WFIRST bins, which leads to a larger number of galaxy-galaxy lensing bins in the data vector.}
\label{fi:LSST_zdistri}
\end{figure}

Adopting these depth cuts and the associated galaxy selection function for source and lens samples we model the true redshift distribution $n_{\mathrm{x}}(z)$ for $\mathrm{x}\in\{\mathrm{lens},\; \mathrm{source}\}$ as 
\be 
\label{eq:redshiftben}
n_{\mathrm{x}}(z) \equiv \frac{d^2 N_{\mathrm{x}}}{dzd\Omega_s} = \bar{n}_{\mathrm{x}}\frac{ \Theta(z_{\mathrm{max}}-z)\; z^{2} \exp \left[ - \left(  \frac{z}{z_0} \right)^{\alpha} \right]}{\int_0^{z_{\mathrm{max}}} dz\;z^{2} \exp \left[ - \left(  \frac{z}{z_0} \right)^{\alpha} \right]} \,.
\ee 
$N_{\mathrm{x}}$ denotes the total number of source/lens galaxies, and $\bar{n}_{\mathrm{x}}$ the effective number density of source/lens galaxies. Two relevant parameters determine the functional form of the distributions, i.e. $(z_0=0.28, \alpha=0.9)$ for the lens sample and $(z_0=0.11, \alpha=0.68)$ for the source sample. 

We impose a high-$z$ cut $z_{\mathrm{max}}=3.5$ for sources and $z_{\mathrm{max}}=1.2$ for lenses, again following the analysis choices of D18, which yield the following number densities for lenses and sources  
\bea
\label{eq:nbarsource}
\bar{n}_{\mathrm{source}} = N_{\mathrm{source}}/\Omega_{\mathrm{s}} &=& 27\;\mathrm{galaxies/arcmin^2} \,, \\ 
\label{eq:nbarlens}
\bar{n}_{\mathrm{lens}} = N_{\mathrm{lens}}/\Omega_{\mathrm{s}} &=& 48\;\mathrm{galaxies/arcmin^2}\,.
\eea
The lens galaxies correspond to the LSST gold sample, which D18 obtain by measuring the number density as a function of magnitude in deep field i-band HSC data \citet{aab17}, fitting it with a power-law to extrapolate to fainter magnitudes, and removing 12\% of the area due to masking. The source galaxy sample in D18 is based on processing image simulations generated with the WeakLensingDeblending software package\footnote{https://github.com/LSSTDESC/WeakLensingDeblending} using an LSST CatSim catalog as the input extragalactic catalog; we refer the reader to the appendix of D18 for the exact implementation. 
For both lenses and sources we define 10 tomographic bins. The source redshift bin limits are chosen such that $\bar{n}^i_{\mathrm{source}}=2.7$ galaxies/arcmin$^2$ and for the lens sample we choose 10 equidistant tomographic bins between 0.2-1.2. The latter implies that the lens galaxy number density per bin is defined as 
\be
\label{eq:nbar}
\bar{n}_x^i = \int dz\; n_x^i(z).
\ee

The true redshift distribution in Eq.~(\ref{eq:redshiftben}) is then convolved with a Gaussian photometric redshift uncertainty model to obtain the distribution within tomographic bin $i$
\be
\label{eq:photoz}
n^i_x(z_{\mathrm{ph}}) = \int_{z^i_{\mathrm{min},x}}^{z^i_{\mathrm{max},x}} dz \, n_{\mathrm{x}}(z) \, p^i\left(z_{\mathrm{ph}}|z,x\right)\,,
\ee
where $p\left(z_{\mathrm{ph}}|z,x\right)$ is the probability distribution of $z_{\mathrm{ph}}$ at given true redshift $z$ for galaxies from population $x$ 
\be
\label{eq:redbin}
p^i\left(z_{\mathrm{ph}}|z,x\right) = \frac{1}{\sqrt{2\pi}\sigma_{z,x}(1+z)}
\exp\left[-\frac{\left(z-z_{\mathrm{ph}} - \Delta^i_{z,x}\right)^2}{2\left(\sigma_{z,x}(1+z)\right)^2}\right]\,.
\ee
The resulting Gaussian tomographic bin is parametrized through scatter $\sigma_z(z)$ and bias between $z-z_{\mathrm{ph}}$, i.e. $\Delta^i_z(z)$. The bias $\Delta^i_z(z)$ has fiducial value of zero; for the lens sample the fiducial $\sigma_z=0.03$ and for the sources the corresponding value is $\sigma_z=0.05$. The resulting distributions are shown in Fig.~\ref{fi:LSST_zdistri}. We detail our modeling of uncertainties in these redshift distribution in Sect.~\ref{sec:sys}.

\subsection{WFIRST} 
\label{sec:HLS}
\begin{figure}
\includegraphics[width=8.1cm]{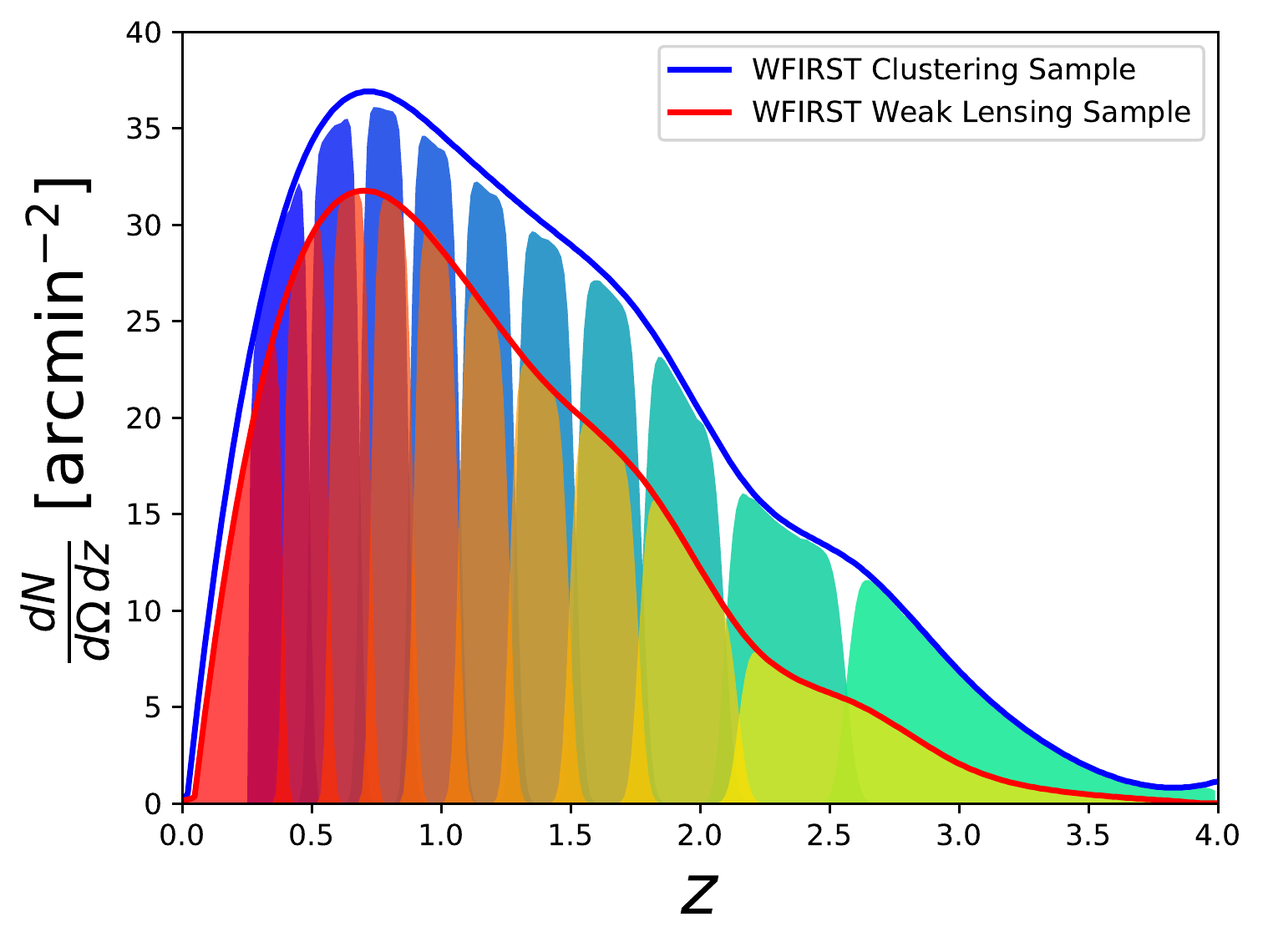}
\caption{This plot shows the WFIRST redshift distribution with the 10 tomographic lens bins (blue) and 10 tomographic source bins (red). The Gaussian photo-z tomographic bin-width ($\sigma_z=0.01$) corresponds to the optimistic scenario that we assume for the combined WFIRST HLS and LSST band coverage, i.e. 4 NIR and 6 optical bands.}
\label{fi:WF_zdistri}
\end{figure}

We use the WFIRST exposure time calculator (ETC) \citep{hgk12} to compute realistic survey scenarios for WFIRST's coverage of area and depth in a given band. We fix the time per exposure and vary the number of exposures to build up depth over the survey area of a given scenario. For the HLS Reference survey this area is 2,000 deg$^2$, for the WFIRST wide scenario it corresponds to the LSST area of 18,000 deg$^2$. The total survey time for a given number of exposures includes a simple prescription for overheads and is correct to approximately 10$\%$. 

In order to obtain accurate redshift distributions we closely follow \cite{hcm19} in applying the ETC results to the CANDELS data set \citep{Candels,candels2}, which is the only data set available that is sufficiently deep in the near-infrared to model WFIRST observations. The ETC has a built-in option to obtain a weak lensing catalog based on an input catalog of detected sources. The criteria for galaxies to be considered suitable for weak lensing are S/N>18 (J+H band combined, matched filter), resolution factor R>0.4, and ellipticity dispersion $\sigma_\epsilon<0.2$, where we use the \cite{bej02} convention (i.e. $\epsilon=(a^2-b^2)/(a^2+b^2)$ instead of $(a-b)/(a+b)$). 
We apply these selections to the CANDELS catalog and obtain our source sample for the WFIRST HLS 4 NIR band survey. For the lens sample we select CANDELS galaxies with S/N$>$10 in each of the 4 WFIRST bands. Our WFIRST analysis assumes LSST photometry from the ground, hence we further down-select both samples by imposing a S/N>5 cut in each LSST band, except for the u-band. We note that nevertheless more than 50\% of our galaxies have S/N>5 in u-band as well.  

The resulting number densities for the HLS are 
\bea
\label{eq:nbarHLS}
\bar{n}_{\mathrm{source}} = N_{\mathrm{source}}/\Omega_{\mathrm{s}} &=& 51\;\mathrm{galaxies/arcmin^2} \\ \,
\bar{n}_{\mathrm{lens}} = N_{\mathrm{lens}}/\Omega_{\mathrm{s}} &=& 66\;\mathrm{galaxies/arcmin^2}\,.
\eea
where $\Omega_{\mathrm{s}}$ is the WFIRST survey area. We impose a $z_\mr{min} = 0.25$ for the lens sample and define 10 tomographic bins such that $\bar{n}^i_{x}=\bar{n}_{x}/10$. These tomographic bins are then convolved with a Gaussian distribution (see Eq. (\ref{eq:photoz})) with mean zero and $\sigma_z=0.01$. This optimistic, narrow width of the Gaussian kernel is motivated by the fact that our HLS survey assumes a joint WFIRST and LSST sample, which altogether gives good photometry in 10 bands ranging from 0.3-2 microns. The resulting redshift distributions are depicted in Fig. \ref{fi:WF_zdistri}.

\subsection{WFIRST Survey Strategy Variations} 
\label{sec:WFIRST_wide}
\begin{figure}
  \includegraphics[width=8.3cm]{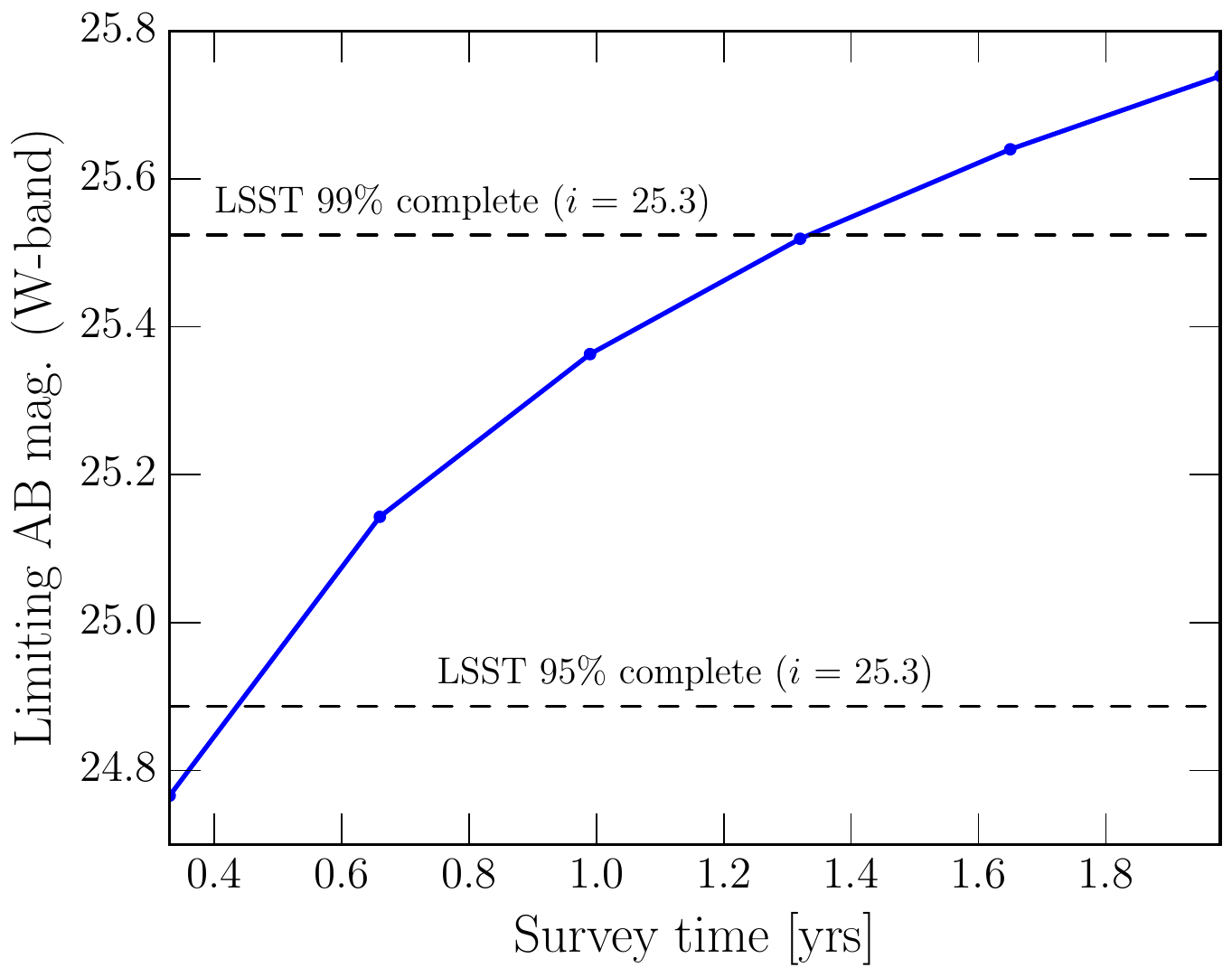}
  \caption{Limiting magnitude of a 18,000 deg$^2$ WFIRST W-band survey as a function of survey time. We also show the 95$\%$ and 99$\%$ completeness thresholds of the LSST gold sample \citep[$i<25.3$, as defined in][, which corresponds to S/N>20 for point sources]{LSST2009} as dashed lines. } 
\label{fi:wbanddepth}
\end{figure}

After defining the LSST and the Reference WFIRST HLS scenario in the past sections we now explore and motivate possible variations in the WFIRST survey strategy, in particular a WFIRST wide scenario that covers the LSST footprint in the W-band (see Table \ref{tab:specs}). We again use the CANDELS catalog when defining galaxy samples for WFIRST and LSST below. 

Figure \ref{fi:wbanddepth} shows the results when using the ETC to compute the depth of the W-band as a function of time under the assumption that no other bands are used. We find that a $\sim$5 month WFIRST W-band survey can obtain high-resolution space imaging for $\sim$95$\%$ of the LSST gold sample (see Fig. \ref{fi:wbanddepth}). 

As a first result of this paper we conclude that if blending poses a systematics limitation to LSST weak lensing cosmology, a dedicated 5 month WFIRST survey would identify almost all LSST blends and enable improved modeling of shapes and photo-z for said blends. 

A 1.3 year WFIRST W-band survey will provide corresponding information for $\sim$99$\%$ of the LSST gold sample and of course also substantially increase the depth of the WFIRST imaging. This opens up the idea to use the deeper WFIRST imaging for shape measurements and combine these with the ground based LSST photometry. 

To explore this idea further we define a WFIRST wide scenario, perform a full simulated likelihood analysis, and compare the results to the constraining power of an LSST Year 10 survey and the Reference WFIRST HLS survey. We assume a 1.5 year WFIRST wide survey in the W-band and follow the same procedure as for the WFIRST HLS survey (Sect.~\ref{sec:HLS}) in deriving the lens and source sample. 

\begin{figure}
\includegraphics[width=8cm]{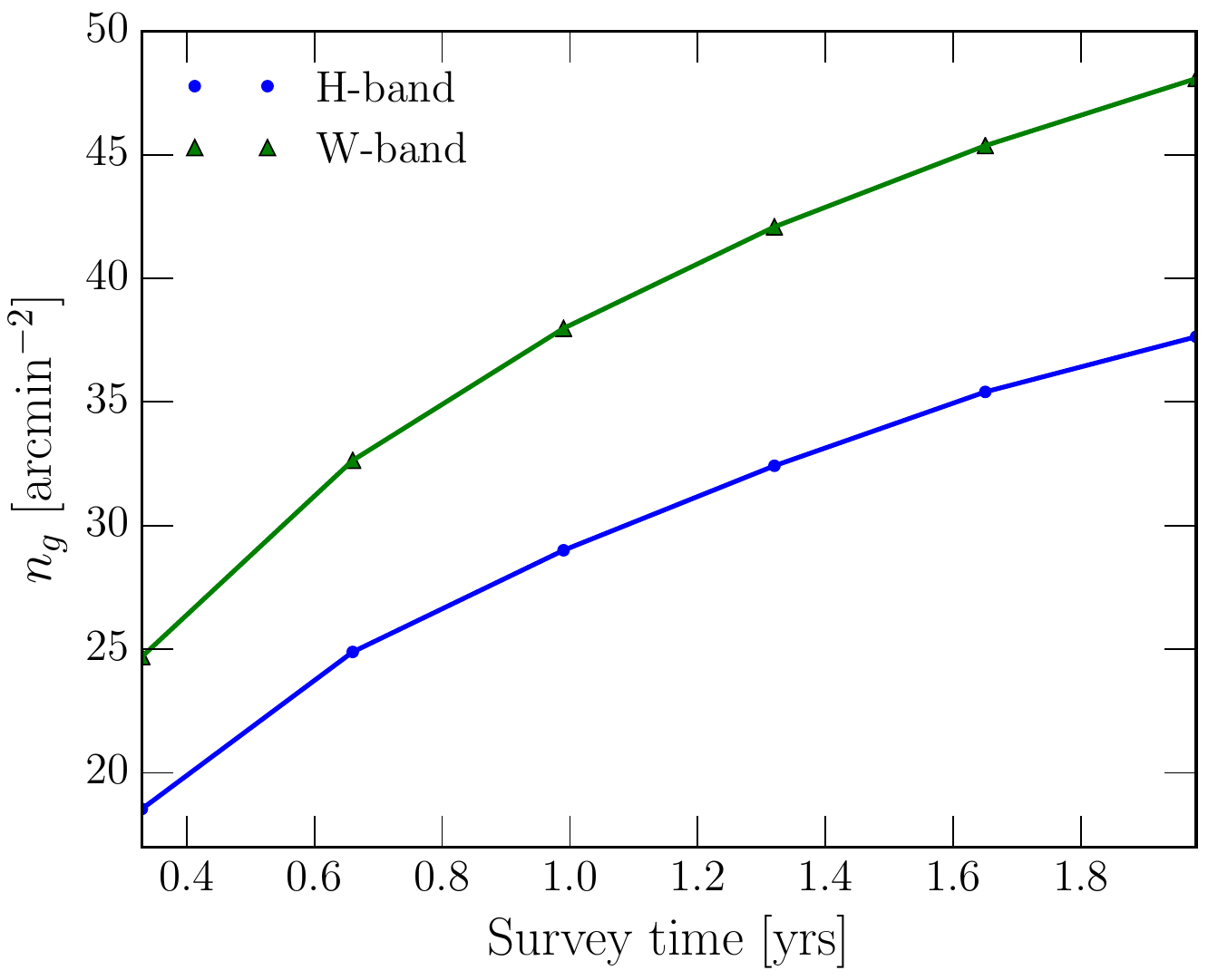}
  \caption{The number density of a WFIRST weak lensing galaxy sample for a 18,000 deg$^2$ survey when conducted in W or H-band, respectively, again as a function of survey time.}
\label{fi:widenumberdensity}
\end{figure}

\begin{figure}
 \includegraphics[width=8cm]{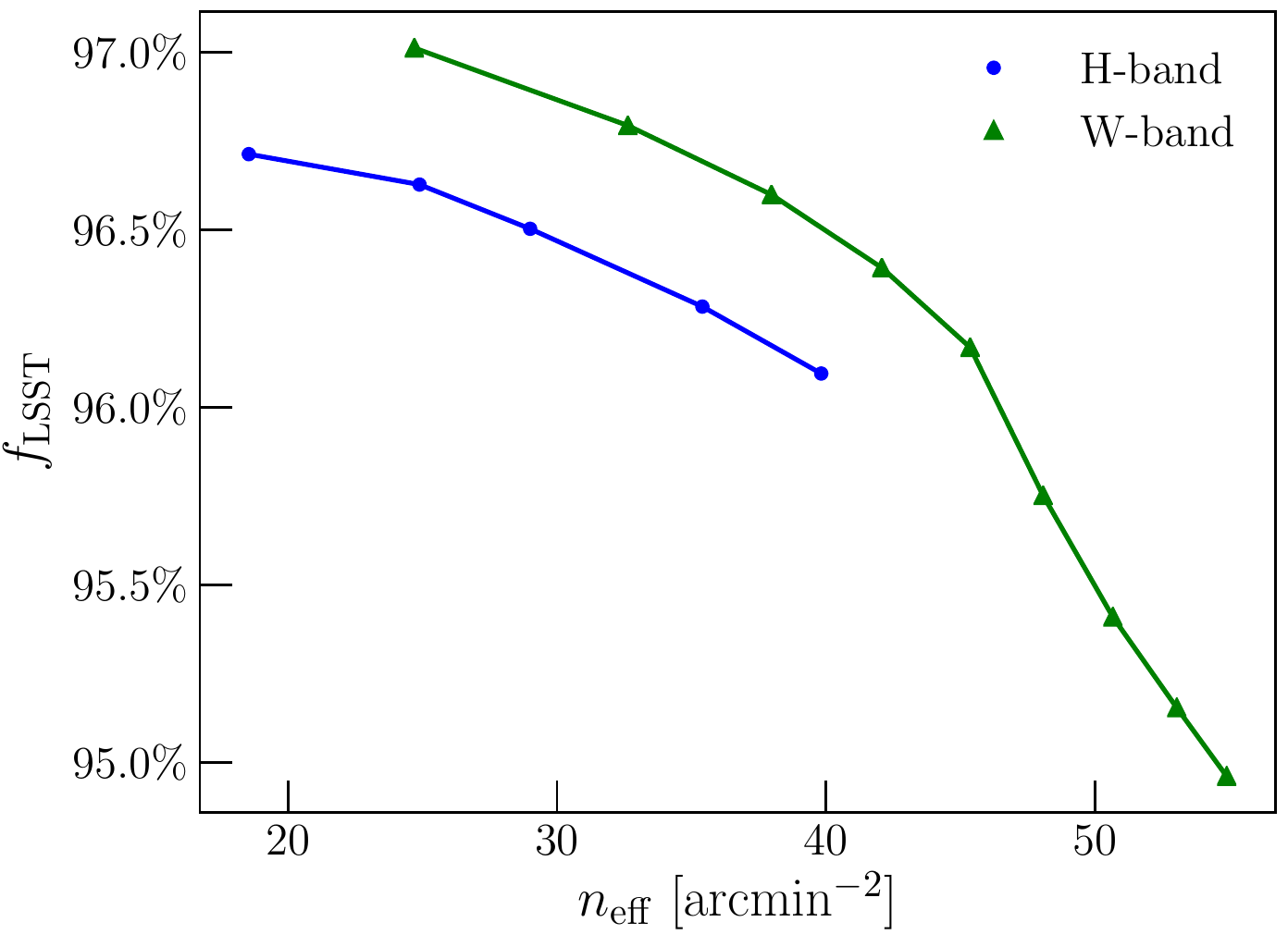}
\caption{Fraction of LSST galaxies with acceptable multi-band photometry as a function of number density of a WFIRST weak lensing sample, based on the CANDELS catalog.}
\label{fi:contour_stats}
\end{figure}

Figure \ref{fi:widenumberdensity} shows the number density of galaxies suitable for shape measurements from a WFIRST 18,000 deg$^2$ survey as a function of survey time and Fig.~\ref{fi:contour_stats} shows the corresponding fraction of LSST galaxies for which good photo-z information (5$ \sigma$ detection in the LSST bands, except for u-band) can be obtained. We also show the corresponding results for the WFIRST H-band, which is useful as an alternative to the W-band since wavelength-dependent Point-Spread Function (PSF) modeling might prohibit shape measurements from a band as broad as the W-band. WFIRST's diffraction-limited PSF size ranges from 0.085\arcsec to 0.175\arcsec over the W-band, which is about a 50\% change, compared to only a 20\% change when using WFIRST's H-band. 

We note that the ESA/NASA Euclid satellite mission is developing mitigation techniques for a similar problem given that Euclid's diffraction-limited PSF size ranges from 0.085\arcsec to 0.155\arcsec over the VIS band, i.e., the main band in which Euclid measures shapes. \cite{cav10, csl18, erh19} propose a variety of methods on how to control wavelength-dependent PSF uncertainties through a combination of improved galaxy spectral energy templates or PSF measurements based on stars that span the same color range as the galaxies. Additional photometric information from the ground is the main avenue for Euclid to gain the relevant information to mitigate this effect, albeit this is of course limited by the difference in resolution of space- and ground-based imaging \citep[also see][for additional wavelength-dependent PSF effects from the atmosphere]{meb15}. 

WFIRST is in the unique position to collect narrow band, space resolution imaging over a smaller, but representative area and calibrate its wide W-band survey if this effect becomes the dominant systematic.  

A 1.5 year WFIRST wide survey would yield 45 galaxies/arcmin$^2$ for the source sample (cf. Fig. \ref{fi:widenumberdensity}) and 68 galaxies/arcmin$^2$ for the lens sample, which is again defined as a S/N>10 cut based on the CANDELS catalog. Since we require good LSST photometry for our WFIRST galaxy sample these number densities are further reduced to 43 galaxies/arcmin$^2$ (cf. Fig. \ref{fi:contour_stats}) for the joint source and 50 galaxies/arcmin$^2$ for the joint lens sample.

The calculation of the redshift distributions follow the same procedure as for the WFIRST HLS survey (Sect. \ref{sec:HLS}), the only difference being that we assume a slightly wider Gaussian kernel $\sigma_z=0.02$ compared to the HLS scenario. 

\section{Likelihood Analysis - Weak Lensing and Galaxy Clustering}

In Sects. \ref{sec:basics} and \ref{sec:surveysims} we describe the basic setup of our analysis including covariance computation, modeling of observables, inference process, galaxy sample selection, and redshift distribution computation. In the following we detail the analysis choices for our likelihood analysis including details on the systematics implementation. 

\label{sec:like}
\renewcommand{\arraystretch}{1.3}
\begin{table*}
\caption{Fiducial parameters, flat priors (min, max), and Gaussian priors ($\mu$, $\sigma$) for WFIRST HLS, LSST and WFIRST wide. We include survey specific parameters with the fiducial values and priors in the upper half of the table, where Gauss($\mu$, $\sigma$) stands for Gaussian priors with mean $\mu$ and deviation $\sigma$. In the lower half we list the cosmological and astrophysical parameters. Flat (min, max) stands for a flat prior between (min, max).}
\begin{center}
\begin{tabular*}{\textwidth}{@{\extracolsep{\fill}}| c c c c c c c |}
\hline
 & \multicolumn{2}{|c}{\tbf{WFIRST HLS}} & \multicolumn{2}{|c}{\tbf{LSST}} & \multicolumn{2}{|c|}{\tbf{WFIRST wide}} \\  
\hline
Parameter & Fid & Prior & Fid & Prior & Fid & Prior \\  
\hline 
\multicolumn{7}{|c|}{\tbf{Survey}} \\
$\Omega_{\mathrm{s}}$ & 2,000 deg$^2$ & fixed & 18,000 deg$^2$ & fixed & 18,000 deg$^2$ & fixed\\
$n_{\mathrm{source}}$ & 51 gal/arcmin$^2$ & fixed & 27 gal/arcmin$^2$ & fixed & 43 gal/arcmin$^2$ & fixed\\
$\sigma_\epsilon$ & 0.26 & fixed & 0.26 & fixed & 0.26 & fixed \\
$n_{\mathrm{lens}}$ & 66 gal/arcmin$^2$ & fixed & 48 gal/arcmin$^2$ & fixed & 50 gal/arcmin$^2$ & fixed\\
\hline
\multicolumn{7}{|c|}{\tbf{Lens photo-z }} \\
$\Delta_\mr{z,lens}^i $ & 0.0 &  Gauss (0.0, 0.001)  & 0.0 &  Gauss (0.0, 0.001)  & 0.0 &  Gauss (0.0, 0.001) \\
$\sigma_\mr{z,lens} $ & 0.01 & Gauss (0.01, 0.002) & 0.03 & Gauss (0.03, 0.003) & 0.02 & Gauss (0.02, 0.002) \\
\hline
 \multicolumn{7}{|c|}{\tbf{Source photo-z}} \\
$\Delta_\mr{z,source}^i $ & 0.0 &  Gauss (0.0, 0.001) & 0.0 &  Gauss (0.0, 0.001) & 0.0 &  Gauss (0.0, 0.001) \\
$\sigma_\mr{z,source}$ &0.01 & Gauss (0.01, 0.002) &0.05 & Gauss (0.05, 0.003) &0.05 & Gauss (0.02, 0.002) \\
\hline
 \multicolumn{7}{|c|}{\tbf{Shear calibration}} \\
$m_i $ & 0.0 & Gauss (0.0, 0.002) & 0.0 & Gauss (0.0, 0.003) & 0.0 & Gauss (0.0, 0.002)\\
\hline 
\hline 
 \multicolumn{7}{|c|}{\tbf{ Common assumptions across all surveys}} \\
\hline
Parameter & \multicolumn{3}{c}{Fid} & \multicolumn{3}{c|}{Prior}  \\  
\hline
\multicolumn{7}{|c|}{\tbf{Cosmology}} \\
$\om$ & \multicolumn{3}{c}{0.3156}&  \multicolumn{3}{c|}{flat (0.1, 0.6)} \\ 
$\sig$ & \multicolumn{3}{c}{0.831} & \multicolumn{3}{c|}{flat (0.6, 0.95)} \\ 
$\ns$ &\multicolumn{3}{c}{ 0.9645} & \multicolumn{3}{c|}{flat (0.85, 1.06)} \\
$\w$ & \multicolumn{3}{c}{ -1.0} &   \multicolumn{3}{c|}{flat (-2.0, 0.0)} \\
$\wa$ &\multicolumn{3}{c}{ 0.0} &  \multicolumn{3}{c|}{flat (-2.5, 2.5)} \\
$\omb$ & \multicolumn{3}{c}{ 0.0492} &  \multicolumn{3}{c|}{flat (0.04, 0.055)}  \\
$h_0$  &\multicolumn{3}{c}{0.6727} &  \multicolumn{3}{c|}{flat (0.6, 0.76) } \\
\hline
 \multicolumn{7}{|c|}{\tbf{Galaxy bias (tomographic bins)}} \\
$b_\mr{g}^i$ & \multicolumn{3}{c}{1.3 + i $\times$ 0.1}  &  \multicolumn{3}{c|}{flat (0.8, 3.0)}\\
\hline
\multicolumn{7}{|c|}{\tbf{Intrinsic alignment}} \\
$A_\mr{IA} $ & \multicolumn{3}{c}{ 5.95} &  \multicolumn{3}{c|}{flat (0.0, 10)} \\
$ \beta_\mr{IA} $ & \multicolumn{3}{c}{1.1} & \multicolumn{3}{c|}{flat (-4.0, 6.0)} \\
$\eta_\mr{IA} $ & \multicolumn{3}{c}{ 0.49} & \multicolumn{3}{c|}{flat (-10.0, 10.0) }\\
$\eta^\mr{high-z}_\mr{IA} $ & \multicolumn{3}{c}{0.0} & \multicolumn{3}{c|}{flat (-1.0, 1.0) }\\
\hline
\multicolumn{7}{|c|}{\tbf{Baryonic Physics}} \\
$Q_1 $ & \multicolumn{3}{c}{ 0.0 } & \multicolumn{3}{c|}{Gauss (0.0, 16.0)} \\
$Q_2 $ & \multicolumn{3}{c}{ 0.0 } & \multicolumn{3}{c|}{Gauss (0.0, 5.0)} \\
$Q_3 $ & \multicolumn{3}{c}{ 0.0 } & \multicolumn{3}{c|}{Gauss (0.0, 0.8)} \\
\hline
\end{tabular*}
\end{center}
\label{tab:params}
\end{table*}
\renewcommand{\arraystretch}{1.0}

\subsection{Building a 3x2pt data vector}
\label{sec:datav}
\paragraph*{Source galaxies -- cosmic shear}
Given the 10 tomographic bins for the source sample we compute 55 cosmic shear auto-and cross power spectra, which we divide into 15 logarithmically spaced Fourier mode bins ranging from $l_{\rm{min}} = 30$ to $l_{\rm{max}} = 3000$. 

\paragraph*{Lens galaxies -- galaxy clustering}
The galaxy clustering data vector is also divided into 15 $l$-bins ranging from 30-3000, however we exclude high $l-$bins, if scales below $R_\mr{\min}=2\pi/k_\mr{max}=21 \mr{\;Mpc/h}$ contribute to the projected integral (see Eq. (\ref{eq:projected})).  

\paragraph*{Lens $\times$ source galaxies -- galaxy-galaxy lensing}
The galaxy-galaxy lensing part of the data vector assumes the lens galaxy sample as foreground and the source galaxy sample as background galaxies; we only consider non-overlapping source and lens in redshift bins. We again impose a cut-off at $R_\mr{\min}=21 \mr{\;Mpc/h}$ for the baseline model. We note that the low, narrow redshift distribution of the 10 lens tomographic bins of the LSST scenario leads to a substantially higher number of galaxy-galaxy lensing power spectra (52), compared to both WFIRST scenarios (32).

\subsection{Systematics} 
\label{sec:sys}
We parameterize uncertainties arising from systematics through nuisance parameters, which are summarized with their fiducial values and priors in Table \ref{tab:params}. Our default likelihood analysis includes the following systematics:

\paragraph*{Photometric redshift uncertainties} As described in detail in Sect. \ref{sec:surveysims} we consider Gaussian photometric redshift uncertainties, which are characterized by scatter $\sigma_z(z)$ and bias $\Delta_z(z)$. While these may in general be arbitrary functions, we further assume that the scatter can be described by the simple redshift scaling $\sigma_{z,x}(1+z)$, and we allow $\sigma_{z,x}$ to vary around its fiducial value. Furthermore, we implement one (constant) bias parameter $\Delta^i_{z,x}$ per redshift bin (cf. Eq. (\ref{eq:redbin})) as a free parameter that is again allowed to vary with Gaussian priors. 

Since we have 10 tomographic bins for the lens and source sample, we vary 20 photo-z bias parameters $\Delta^i_{z,x}$ and two photo-z scatter parameters $\sigma_{z,x}(1+z)$. We note that the selection and characterization of lens and source galaxy samples are some of the most challenging choices in a multi-probe analysis and variations warrant future investigation. The trade space of photo-z accuracy versus number density, e.g., by using a ``redmagic'' sample \citep{rra15} instead of the LSST Gold sample is interesting to explore further. 

\paragraph*{Linear galaxy bias} is described by one nuisance parameter per lens galaxy redshift bin $b_\mr{g}^i$, which is marginalized over using conservative flat priors $[0.8;3.0]$. 

We note that the fiducial values of the galaxy bias parameters have little impact on the results, however, a more stringent prior would be highly beneficial. For example, the interplay of galaxy bias and photo-z uncertainties limits the ability of galaxy-galaxy lensing to self-calibrate intrinsic alignment models. Prior information on linear galaxy bias parameters would also help implement higher order bias models that require additional free parameters but that would allow to push to smaller scales in clustering and galaxy-galaxy lensing \citep{dds18,isz19}. 

\paragraph*{Shape measurement uncertainties} are a primary concern for all weak lensing based cosmology endeavors. Substantial progress has been made in the past years to model and control shape measurement uncertainties \citep{shh17,hum17}. For LSST the atmospheric PSF and blended objects remain major obstacles given the high level of required precision \citep{dst16,mmj18}. For space-based missions we already mentioned the wavelength dependent PSF as a major uncertainty (see Sect. \ref{sec:WFIRST_wide}); since WFIRST will use H4RG-10 infrared detectors, nonlinear detector effects such as the brighter fatter effect and nonlinear inter-pixel capacitance will need to be fully characterized before launch \citep{pss18,hic19,chh19,fgc20}.  \citet{chh19} measured these nonlinearities for a prototype detector via a correlation analysis of flat field data, with a statistical precision that meets WFIRST requirements.  Further laboratory studies of these nonlinear detector effects are underway that will bolster confidence in our ability to accurately calibrate galaxy shapes.  

In this paper we model residual shape measurement uncertainties as multiplicative shear calibration, specifically, we use one parameter $m^i$ per redshift bin, which affects cosmic shear and galaxy-galaxy lensing power spectra as
\bea
\nonumber C_{\kappa \kappa}^{ij}(l) \quad &\longrightarrow& \quad (1+m^i) \, (1+m^j) \, C_{\kappa \kappa}^{ij}(l), \\
C_{\delta_{\mathrm{g}}\kappa}^{ij}(l) \quad &\longrightarrow& \quad (1+m^j) \, C_{\delta_{\mathrm{g}}\kappa}^{ij}(l),
\eea
The fiducial value of each $m^i$ is zero and we marginalize over the $m^i$ independently with Gaussian priors of different width for LSST and WFIRST (LSST priors are twice as large as WFIRST priors). Altogether shape uncertainties add 10 nuisance parameters to our likelihood analyses.

\paragraph*{Intrinsic alignment} 
Intrinsic alignment (IA) of source galaxies has been studied extensively as a systematic for weak lensing through observations, simulations, and theory \citep[e.g.,][]{his04,mhi06, job10, smm14, tri14,tsm15, bvs15, ccl15, vcs19} 

The assumption that the shape and orientation of an elliptical galaxy are determined by the shape of the halo in which it resides, causes a correlation of intrinsic galaxy ellipticities with the gravitational tidal field. We employ the so-called ``tidal alignment'' model \citep{ckb01}, specifically we use the nonlinear alignment (NLA) version \citep{his04, brk07} of the tidal alignment model to describe IA of elliptical (red) galaxies. This approach captures most to the IA signal and we do not consider quadratic alignment/tidal torquing models \citep[see e.g.][]{bmt19} or more sophisticated IA modeling as a function of galaxy color \citep{sbt19} in this paper.

Our implementation follows \cite{keb16} (K16 hereon) for cosmic shear and \cite{kre17} for the extension to galaxy-galaxy lensing. The cosmic shear and galaxy-galaxy lensing projected power spectra are modified by additive terms to include the IA effects, specifically 
\bea
\label{eq:CgI}
C_{\kappa \kappa}^{ij}(l) &\rightarrow& C_{\kappa \kappa}^{ij}(l) + C_\mathrm{II}^{ij}(l) + C_\mathrm{GI}^{ij}(l) \,, \\
C_{\delta_{\mathrm{g}}\kappa}^{ij}(l) &\rightarrow& C_{\delta_{\mathrm{g}}\kappa}^{ij}(l) + C_{\delta_{\mathrm{g}}\mr I}^{ij}(l) \,.
\eea
The $C_\mathrm{II}^{ij}(l)$ term arises since galaxies that are physically close experience an alignment of their intrinsic ellipticity (II) from the same tidal field. For galaxy pairs that are separated in redshift, alignment effects also occur: foreground galaxies are aligned by the same tidal field that lenses background galaxies, which introduces a (anti-) correlation of shear (G) and intrinsic (I) ellipticity.

The $C_{\delta_{\mathrm{g}}\mr I}^{ij}(l)$ effect describes a correlation of galaxy overdensity and intrinsic ellipticity (I) for galaxy pairs that are physically close and affected by the same tidal field. 

The projected power spectra can be computed from the corresponding 3D power spectra as described in Eq. (\ref{eq:projected}), specifically
\bea
\label{eq:IA1}
C_\mr{II}^{ij} (l) &=&  \int \mr d \chi \, \frac{q^i_{\delta_{\mathrm{g}}} (\chi) q^j_{\delta_{\mathrm{g}}} (\chi)}{\chi^2} \, f_\mr{red}^2(m_{\mathrm{lim}},z)\, P_\mr{I I} (k,z) \,, \\
\label{eq:IA2}
C_\mr{GI}^{ij} (l) &=&  \int \mr d \chi \, \frac{q^i_{\delta_{\mathrm{g}}} (\chi) \,  q^j_\kappa (\chi)}{\chi^2}  \, f_\mr{red}(m_{\mathrm{lim}},z)\, P_\mr{G I} (k,z) \,, \\
\label{eq:IA3}
C_{\delta_{\mathrm{g}}\mr I}^{ij}(l) &=& \int \mr d \chi \, \frac{q^i_{\delta_{\mathrm{g}}} (\chi) \,  q^j_\kappa (\chi)}{\chi^2} f_{\mathrm{red}}(m_{\mathrm{lim}},z) \, P_\mr{\delta_g I}(k,z) \,,
\eea
where $f_{\mr{red}}$, the fraction of red galaxies at redshift $z(\chi)$, is evaluated from the GAMA luminosity function \citep{lnb12} assuming a limiting magnitude $m_{\mathrm{lim}} = 25.3$. 

In the tidal alignment picture, the intrinsic ellipticity field is to leading order linear in the density field and we can write the relevant 3D-IA power spectra as
\bea
P_{\mathrm{II}}(k,z) &=& A^2(m_{\mathrm{lim}}, z) \, P_{\delta \delta}(k,z)\, , \\
P_{\mathrm{G I}}(k,z) &= &-A(m_{\mathrm{lim}}, z) \, P_{\delta \delta}(k,z)\,, \\
\label{eq:IA33}
P_\mr{\delta_g I}(k,z) &=& -A(m_{\mathrm{lim}},z) \, b_g(z) \, P_{\delta \delta}(k,z)\,.
\eea
Uncertainties in galaxy bias in Eq. (\ref{eq:IA33}) are again parameterized as one free parameter per tomographic lens bin.  

Our likelihood analysis includes uncertainties in IA modeling via four parameters that enter the IA amplitude $A(m_{\mathrm{lim}},z)$ (see Table~\ref{tab:params}). The IA amplitude, for a given limiting magnitude, can be computed as a function of redshift and luminosity function of a given galaxy sample (LRGs in our case)
\be
\label{eq:A_L}
A(L,z)  = \frac{C_1\rho_{\rm m,0}}{D(z)}A_0 \left(\frac{L}{L_0}\right)^\beta \left(\frac{1+z}{1+z_0}\right)^{\eta}\,,
\ee
where $C_1 \, \rho_\mr{cr}=0.0134$ is derived from SuperCOSMOS observations \citep{his04, brk07}. As fiducial values (cf. Table \ref{tab:params}) for our nuisance parameters $A_0,\eta,\beta$ we adopt the constraints from the MegaZ-LRG + SDSS LRG sample \citep{jma11} with $z_0 = 0.3$ as the observationally-motivated pivot redshift and $L_0$ as the pivot luminosity corresponding to an absolute $r$-band magnitude of $-22$. 

We compute the IA amplitude at given redshift as the average over the alignment amplitudes of red galaxies using the luminosity distribution of the GAMA survey \citep{lnb12}  
\be
A(m_{\mathrm{lim}},z) = \Big\langle A(L,z)\Big\rangle_{\phi_{\rm{red}}}\times \left[\Theta(z_1 -z)+\Theta(z -z_1) \left(\frac{1+z}{1+z_1}\right)^{\eta_\mr{high-z}}\right]\,,
\ee
where $\Theta$ is the step function, which separates the range where our fiducial redshift scaling is based on the MegaZ-LRG + SDSS LRG sample, which extends to $z  \leq  0.7$. Given the substantially larger range of our galaxy sample we extrapolate this functional form, but introduce additional freedom.  The term in square brackets is a truncated power-law in $(1+z)$ with the additional uncertainty parameterized as $\eta_{\rm{high-z}}$ for $z> z_1=0.7$.  This $\eta_{\rm{high-z}}$ parameter also captures uncertainties in the extrapolation of the GAMA luminosity function to higher redshift.

We do not consider additional uncertainties in the luminosity function \citep[cf.][where said uncertainties are marginalized over 6 additional parameters]{keb16}, but we note that these uncertainties can be significant. This is the interface where cosmology meets galaxy formation and we acknowledge that it is critical to understand the latter to precisely constrain the former.

\paragraph*{Baryonic physics}
Baryonic physics effects on the modeling of small scales in the matter power spectrum \citep[e.g.][]{dsb11,crd18} are a pressing concern for cosmic shear \citep[see e.g.][]{shs11,zsd13,ekd15,hem19} and will become a pressing concern for galaxy-galaxy lensing and galaxy clustering if higher-order bias models are included that allow the inclusion of smaller scales.

We mitigate the impact of baryonic physics in two different ways \citep[cf.][for an overview on mitigation strategies]{cmj19}: First, the scale cuts for the galaxy clustering and galaxy-galaxy lensing part of the data vector that are driven by our assumption of using linear galaxy bias, i.e. $R_\mr{\min}=21 \mr{ Mpc/h}$ are conservative and probably sufficient to control baryonic effects for these parts of the data vector. For cosmic shear, our $l_{\rm{max}} = 3000$ cut also mitigates the impact of baryonic physics but as \citet[][H19 from hereon]{hem19} have shown this is insufficient as a scale cut to control baryons at the LSST Year 10 level. 

We employ ``Method C'' detailed in H19 to account for residual uncertainties in baryonic physics in our full 3x2pt data vector. In short, we compute the difference of dark matter only to baryonic 3x2 data vectors for LSST and WFIRST for 5 different baryonic scenarios. The baryonic scenarios are extracted from hydro-simulations, in particular we use the MassiveBlack-II simulation \citep{kdc15},  IllustrisTNG \citep{wsp18,psn18}, Horizon-AGN  \citep{dpp16}, Eagle simulation \citep{scb15} , and the OWLS AGN simulation \citep{sdb10,dsb11}.  
This choice of simulations is motivated by the fact that we require AGN feedback to be included and that we require sufficiently high resolution to reliably model small scales (again see H19 for a summary of the simulations and motivation of this choice). 

From the 5 baryonic scenarios we extract 3x2pt data vectors with the exact properties as discussed in Sect.~\ref{sec:datav} and we build a ``difference matrix''  of baryonic and dark matter data vectors (see Eq. (20) in H19) 
\be 
\label{eq:DiffMatrix_chy}
\begin{aligned}
{\bm \Delta}(\pco) &= 
\left[
  \begin{array}{cccc}
    \vertbar  &        & \vertbar \\
    \B_{\rm 1}-\M    & \ldots & \B_{\rm 5}-\M_{\rm}    \\
    \vertbar  &        & \vertbar 
  \end{array}
\right]_{N_{\rm data} \times N_{\rm sim}} \\
 \end{aligned}
\ee 
We weigh this difference matrix with respect to the statistical uncertainty described in our covariance matrix $\C$, i.e. we perform a Cholesky decomposition $\C=\L \L^{\rm t}$ and compute 
\be 
 {\bm \Delta_{\rm ch}} =\invL {\bm \Delta} = \U_{\rm ch} \ \bm \Sigma_{\rm ch}\ \V_{\rm ch}^{\rm t} \,,
\ee 
where in the last step we perform a singular value decomposition on the weighted difference matrix to extract the Principal Components (PCs) that span the range of uncertainty in baryonic physics. 
The first 5 columns of the $\U_{\rm ch}$ matrix form a complete  description of baryonic uncertainties given our 5 input hydrodynamical scenarios $\B_i$.
\be \label{eq:PCspan_ch}
{\invL (\B_i - \M)  =  \sum_{n = 1}^{5} Q_n\ {\mathbf {PC}}_n \ .}
\ee
By reorganizing Eq. (\ref{eq:PCspan_ch}), we can generate model vectors that include dark matter and baryonic physics as 
\be 
\label{eq:MC}
\M(\pco, {\mathbf Q} ) = \M(\pco) + L \sum_{n = 1}^{m} Q_n\ {\mathbf {PC}}_n \ , 
\ee
where $m \leqq 5$. We include baryons in our analysis by replacing $\M$ in Eq. (\ref{eq:like}) with Eq. (\ref{eq:MC}). We note that the cosmology dependence only enters through the dark matter part of the model vector, while the amplitudes of PCs are used as higher order correction for baryonic effects. We include the first 3 PCs in our likelihood analysis and consequently marginalize over 3 PC amplitudes ($Q_{1\sim3}$) as additional degrees of freedom to model baryonic physics.   

The priors for the $Q$ are highly conservative and chosen such that the 1$\sigma$ region of the Gaussian prior corresponds to twice the amplitude of $Q$'s needed to capture the Illustris (not TNG) simulation. The original Illustris simulation has a very strong feedback scenario which is already highly unlikely given present observations. 

\subsection{Simulated likelihood analysis}
Using the data vectors defined in Sect. \ref{sec:datav} and the analysis choices defined in Table \ref{tab:params}, we simulate likelihood analyses for the LSST Year 10, WFIRST HLS, and WFIRST wide scenarios. We point out that the latter two assume that LSST data exists over the corresponding area. Our likelihood analyses span 56 dimensions, 7 of which relate to cosmological parameters and the remaining 49 describe uncertainties in modeling observational (shear calibration, lens and source photo-z uncertainties) and astrophysical systematics (galaxy bias, intrinsic alignment, baryonic physics). 

\begin{figure}
 \includegraphics[width=8cm]{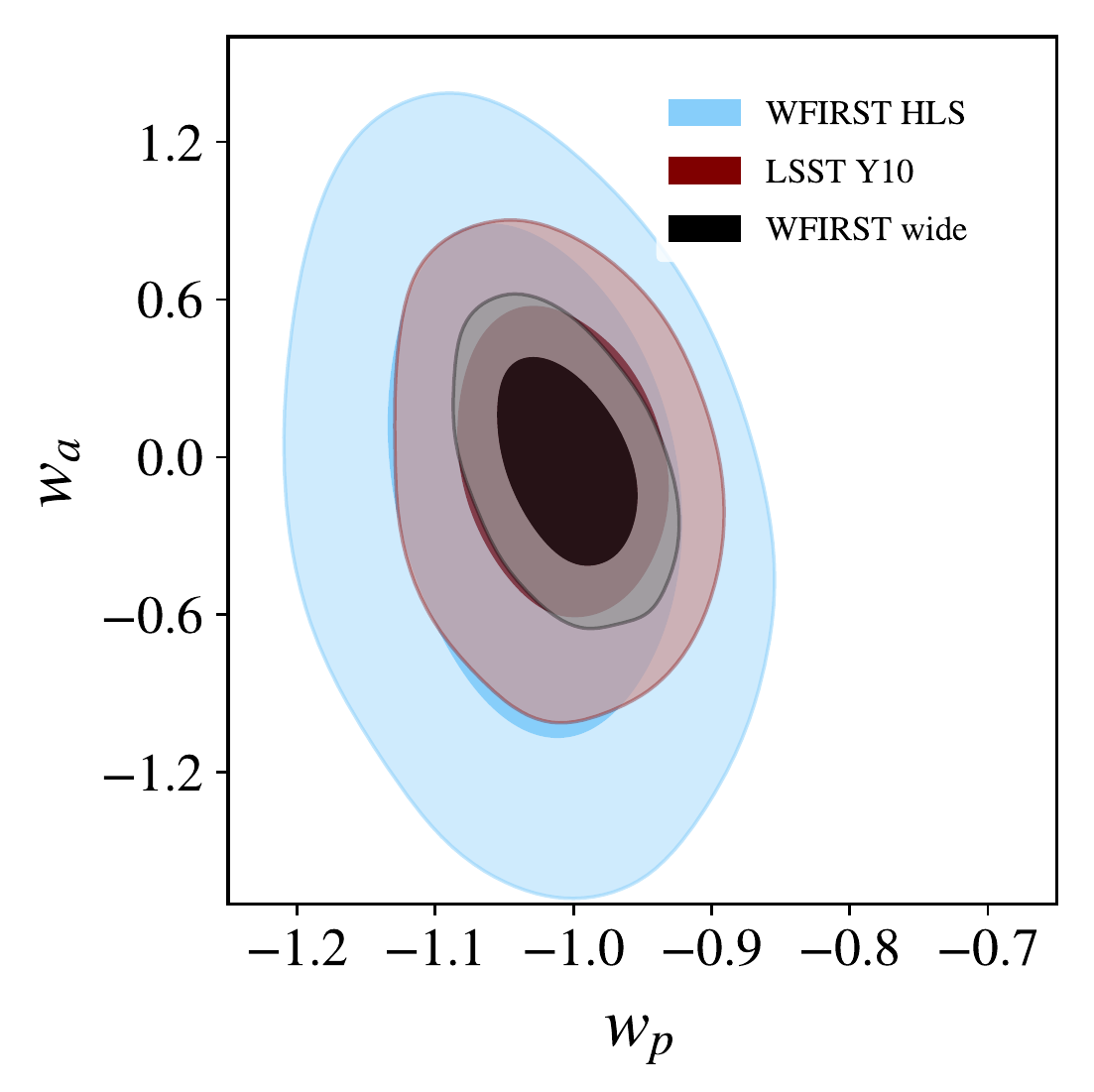}
\caption{
Constraining power on dark energy parameters $w_p$ (where $p$ is a pivot redshift, chosen such that the two parameters are somewhat de-correlated) and $w_a$, marginalized over 5 other cosmological parameters and 49 nuisance parameters describing observational and astrophysical systematics. We show results for the 18,000 deg$^2$ LSST Year 10 data set in \textit{red}, for the 2000 deg$^2$ WFIRST HLS survey in \textit{blue}, and for the WFIRST wide survey (\textit{black}). Both WFIRST scenarios assume LSST multi-band photometry over the corresponding area. We show the 68\% and 95\% contours. 
} 
\label{fi:WF_LSST}
\end{figure}

We use the fast $\CL$ forecasting modules and the \texttt{emcee} parallel sampling algorithm to generate chains with 1120 walkers, 8000 steps each. Altogether our chains comprise 8.96M steps. We compare constraining power of the different scenarios through contour plots in the dark energy equation of state parameters (see Fig. \ref{fi:WF_LSST}) and by computing the Dark Energy Task Force Figure of Merit, $\mr{FoM}=|\C^{-1}_{w_0,w_a}|^{1/2}$. In other words, we obtain the FoM for a given scenario from the MCMC chains by computing the parameter covariance matrix, extracting the $w_0, w_a$ submatrix, inverting it, and computing the square root of its determinant. 

To test sampling convergence, we compare sub-chains of 140,000 steps starting at step 2.1M and find that our FoMs have stabilized at 6M steps. We also vary the number of walkers and starting points (including their variance) and found our results independent of reasonable choices in these settings.

Figure \ref{fi:WF_LSST} shows the constraining power in the dark energy equation of state parameters $w_a$ and $w_p$. The latter corresponds to $w_0$ but computed at a different redshift, here  $z_\mr p=0.4$, to de-correlate the two parameters and to enable the reader to better estimate 1D projected error bars. The contours show a substantial increase in the ability to constrain time-dependent dark energy for the WFIRST wide scenario compared to the LSST Y10 scenario and compared to the WFIRST HLS survey.  

Regarding the WFIRST HLS versus WFIRST wide comparison (blue versus black contours in Fig.~\ref{fi:WF_LSST}) the gain in constraining power (FoM$_\mr{Wwide}$= 5.5 FoM$_\mr{HLS}$)

is mostly driven by the increased area 18,000 deg$^2$ versus 2,000 deg$^2$. The larger number density of lens and source galaxies and the improved photo-z accuracy of the HLS cannot compensate for this effect. It is interesting to see that at the precision of WFIRST HLS and WFIRST wide the results are not fully systematics dominated but that an increase in area has a substantial impact on the contours. 

When comparing LSST to WFIRST wide (red versus black contours in Fig.~\ref{fi:WF_LSST}) the increase in constraining power (FoM$_\mr{Wwide}$= 2.4 FoM$_\mr{LSST}$) stems from a combination of increase in number density, deeper redshift distributions, and improved control of systematics, most importantly photo-z calibration.   
A more detailed study is required to evaluate the exact trade space, in particular since photo-z parameters are highly correlated with our ability to model galaxy bias and intrinsic alignment. Such a study should also include more realistic photo-z errors, such as catastrophic outliers and it should investigate higher-order bias models to push to smaller scales in the galaxy-galaxy lensing and galaxy clustering parts of the data vector. 

It is interesting to compare the LSST and WFIRST HLS contours (blue versus red contours in Fig. \ref{fi:WF_LSST}) given that these differ most significantly in terms of assumptions that enter the likelihood analysis. The substantial difference in area 18,000 deg$^2$ for LSST versus 2,000 deg$^2$ for WFIRST HLS is countered by a significantly higher number density of galaxies (cf.\ Table \ref{tab:params}), deeper redshift distribution, higher photo-z and shape measurement accuracy for the WFIRST HLS scenario. Despite these improvements the increase in area is more important and favors the LSST only scenario, albeit not at a level one might have suspected. A simple 1/area scaling would suggest a factor of 9 improvement in constraining power, however the difference between HLS and LSST Y10 is FoM$_\mr{Wwide}$= 2.4 FoM$_\mr{LSST}$. We attribute this difference in part to the larger number density of WFIRST, but also to the improved systematics control (width of the photo-z bins and priors on how well we know this width and shear calibration).

\section{Discussion}
\label{sec:discussion}
In terms of how to fit a wide survey into the WFIRST mission, several options arise: 
\begin{enumerate}
\item WFIRST does not have expendables that would prohibit extended observations beyond its 5-year primary mission (WFIRST carries enough propellant for at least 10 years of observations, and there are no active cryogens). Dedicating 1.5 years of such an extension to a W-band survey as detailed above, which would be well-matched in terms of timescale to LSST Y10, is a possible scenario. 
\item A second option is to get the wide survey data earlier and to reduce the 2000 deg$^2$ footprint of the HLS in the primary mission. As an extreme example one could even fit the 1.5 year W-band survey outlined above into the nominal 1.6 year HLS survey. This however would come at the very high cost of having almost no grism spectroscopy or multi-band WFIRST coverage; given how important corresponding data is for systematics control, this option appears less favorable.
\item A third option would be to cover a subset of the 18,000 deg$^2$ with the W-band in the WFIRST primary mission, e.g. one could survey the 10,400 deg$^2$ of sky with E(B-V)$<$0.08 that pass within 32 degrees of zenith at LSST. This will likely be the best part of the LSST footprint for extragalactic studies, and it can be surveyed in approximately 1 year to the same depth as the 1.5 year survey considered in this paper. 
\end{enumerate}

We again refer to Fig. \ref{fi:wbanddepth}, as one of the main results of our paper, which shows that already a 5-month WFIRST survey in the W-band can provide 18,000 deg$^2$ high-resolution imaging for $>$95\% of the LSST Year 10 gold galaxy sample (S/N>20 for point sources). For the 10,400 deg$^2$ with E(B-V)$<$0.08 this type of survey would only take $\sim$14 weeks. It is important to note that the WFIRST observing time, including the survey design of the HLS, has not been allocated to date, and that corresponding decisions will depend on the science landscape in $\sim$5 years. In particular, it will be important to see how strongly the first 3 years of LSST data are affected by systematics and how much core science interests in the community (not just cosmology) would benefit from rapid WFIRST wide coverage.   

This analysis does not recommend replacing the Reference 2,000 deg$^2$ HLS survey design with a wide W-band survey of WFIRST. The HLS ensures exquisite systematics control and it is the consensus in the community that systematics control will be more important than maximizing statistical power. We instead recommend the exploration of the WFIRST wide survey strategies in combination with the HLS approach, specifically we envision that a wide WFIRST survey component would use the HLS to anchor shear and photo-z calibration. 

\subsection{WFIRST wide synergies beyond weak lensing and galaxy clustering}

A WFIRST wide survey in one band as described above opens numerous science applications in cosmology beyond the joint weak lensing and galaxy clustering dark energy science depicted in Fig. \ref{fi:WF_LSST}.  

\paragraph*{Galaxy cluster science:} An 18,000 deg$^2$ W-band survey would substantially enhance the weak lensing mass calibration of clusters, a key ingredient for cluster cosmology \citep{laa14,lma14,mla15,mbh19,crs19,wws19,sww20}.  Both the increased spatial resolution and the broader wavelength coverage will help to decrease systematic uncertainties due to blending and photo-z misclassification, which are amplified in overdense cluster fields.  An important aspect is that the W-band addition would enable precise mass calibration to higher redshift clusters than LSST alone \citep[cf.][]{sad18}.
\paragraph*{Trough cosmology:} Cosmic underdensities have emerged as an interesting probe of structure formation and hence dark energy and modified gravity \citep{kcd13,mss14,hps16,dcb19}; potentially corresponding observables are easier to model compared to probes that rely on higher-density environments. Trough cosmology \citep{gfa16,gfk18}, and especially trough lensing, would benefit from the 18,000 deg$^2$ wide WFIRST survey scenario because of the higher density of source galaxies (for trough lensing) and the higher density of detected galaxies that signify the trough boundaries. 
\paragraph*{Strong Lensing:} LSST will find a large number of strong lenses \citep{ogm10}. Color and more importantly shape information from overlapping space observations will allow for significantly improved constraints on the value of $H_0$ and other core cosmology questions \citep{btr19} via enhanced control of lens profile modeling uncertainties.

\paragraph*{Stellar Astronomy:}WFIRST's infrared measurements will enhance the analysis of the stellar population in LSST.  WFIRST's infrared observations will be particularly valuable for tracing brown dwarfs and AGB stars. WFIRST should be able to achieve single exposure precision of 0.01 pixels or 1.1 mas \citep{Sanderson2017}.  Thus, with a 1.5 year survey spread out over 5 years of mission time, WFIRST will measure proper motions with uncertainties of $\sim 200 \mu$as/yr for stars.  This will help trace stellar streams in the galactic halo $\sim 5$ magnitude fainter than GAIA.

\subsection{WFIRST wide synergies with CMB surveys}

The new generation of CMB experiments, e.g. the Simons Observatory \citep[SO, ][]{SO19} and CMB-S4 \citep{CMBS4}, will be well underway by the mid-2020s. In particular, the survey of the Large Aperture Telescope of the Simons Observatory over the full LSST footprint will likely be completed or near completion, providing a detailed map of the integrated matter density through CMB lensing, the integrated pressure distribution through the thermal Sunyaev-Zeldovich' (tSZ) effect, the integrated electron momentum distribution through the kinematic Sunyaev-Zeldovich (kSZ) effect, and the cosmic infrared background (CIB). The increased number density of galaxies and precision in shape measurements from space-based imaging from WFIRST over the full LSST area would therefore enable a new level of cross-correlations measurements between galaxy surveys and CMB surveys.

The CMB lensing kernel function,
\be
\label{eq:qkappa_cmb}
q_{\kappa_{\rm CMB}}(\chi) = \frac{3 H_0^2 \Omega_m }{2 \mathrm{c}^2}\frac{\chi}{a(\chi)}  \frac{\chi^*-\chi}{\chi^*} \,
\ee
peaks at $z\sim2$ and is sensitive to large-scale structure between the observer and the last scattering surface (here $\chi^*$ is the comoving distance to this surface).
Measurements of cross-correlations of galaxy clustering and shapes with CMB lensing would strongly benefit from the higher density of galaxies of a joint catalog, which is beneficial to constraints on cosmic acceleration in several ways.
First, extra cosmological information is held in these cross-correlations, which can alleviate parameter degeneracies, for instance between galaxy bias and cosmological parameters. Second, these measurements will enable further cross-calibration of shear multiplicative biases, as shown in \cite{ske17}, and other nuisance parameters such as uncertainties on photometric redshifts. Third, the joint catalog will have higher density of galaxies at high redshifts, which increases the S/N of CMB lensing. The comparison of constraints obtained from subsets of these high-precision cross-correlation measurements will enable various tests of the consistency of the cosmological model, and if they are consistent, it will enable a new level of constraining power on cosmological physics.

By cross-correlating the CMB lensing map with the large-scale structure traced by photometric redshift slices, we will be able to determine the evolution of the amplitude of density fluctuations with redshift, $\sigma_8(z)$. \citet{Krolewski2019} obtain a S/N of 58 measurement with unWISE x Planck.  For the much higher number density WFIRST/LSST sample correlated with the much lower noise SO lensing map, the S/N would be much higher.

Similarly, thermal SZ observations from CMB surveys will detect a wealth of clusters (16,000 clusters forecasted for SO), all of which benefit strongly from cluster mass calibration through high-precision weak lensing from WFIRST \citep[see e.g.][for corresponding applications to the Atacama Cosmology Telescope and the South Pole Telescope]{mbh19,bds19}. The increased accuracy in photometric redshifts of the joint catalog would improve the line of sight localization of these clusters and it would extend the redshift range of clusters for which both CMB and optical/infrared observations are matched. A joint catalog will also enable improved measurements of cluster profiles and features such as the splashback radius observed in density and lensing profiles \citep{sab19}.  The WFIRST lensing measurements could determine the relationship between integrated pressure ($Y$) and mass. This observation would determine the gas fraction in $\sim 10^{14} M_\odot$ groups, which could then calibrate the effect of baryon feedback on the dark matter distribution \citep{vanDaalen2019}.  

The combination of CMB kSZ measurements and WFIRST/LSST can be used to trace the distribution of electrons around galaxies either through cross-correlations with the optical shear measurements \citep{Dore2004} or with the galaxy distribution \citep{Ferraro2016, Hill2016}. By measuring these effects as a function of redshift, this combination can trace the evolution of the circumgalactic medium \citep{battaglia2019}.

\section{Conclusions}
\label{sec:conc}
LSST, SPHEREx, Euclid, WFIRST in combination with other spectroscopic surveys  (DESI, PFS, 4MOST) and CMB surveys (Simons Observatory, CMB-S4) pose exciting opportunities for discovery in cosmological physics. WFIRST will likely be the latest mission to join this ensemble of experiments, but given its versatile observing capabilities and its flexibility in terms of survey strategy, being last is where WFIRST can synergize best.

In this paper we focus on synergies between WFIRST (space-based, high resolution imaging, NIR multi-band photometry, grism spectroscopy) and LSST (6-band photometry in the visible wavelengths, 10 deg$^2$ FoV, rapid and repeated coverage of 18,000 deg$^2$). We explore alternative scenarios to the reference WFIRST HLS, which assumes LSST and WFIRST data over 2000 deg$^2$, in particular we quantify scenarios where WFIRST covers the entire LSST area rapidly in one band. Our survey simulations are based on the WFIRST exposure time calculator and redshift distributions from the CANDELS catalog.

As a first result, we find that already a 5-month WFIRST survey in the W-band can cover the full LSST survey area providing high-resolution imaging for $>$95\% of the LSST Year 10 gold galaxy sample. Given that blending is a potentially limiting issue for LSST cosmology, the concept of a 5-month WFIRST wide survey is an important idea to study further.

If WFIRST were to spend 1.5 years covering the LSST area in the W-band it would be able to provide high-resolution imaging for $>$99\% of the LSST Year 10 gold galaxy sample. For this second scenario we run a full 3x2pt likelihood analysis that includes non-Gaussian covariances, and accounts for observational and astrophysical systematics (shear calibration, lens and source photo-z uncertainties, galaxy bias, intrinsic alignment, baryonic physics). We run similar 3x2pt likelihood analyses for the standard WFIRST HLS survey and for a LSST Year 10 survey in order to compare to the WFIRST wide concept. We find a significant increase in constraining power for the joint LSST+WFIRST wide survey compared to LSST Y10 (FoM$_\mr{Wwide}$= 2.4 FoM$_\mr{LSST}$) and compared to WFIRST HLS (FoM$_\mr{Wwide}$= 5.5 FoM$_\mr{HLS}$). 

The assumed survey area in both the 5 month and the 1.5 year LSST+WFIRST wide survey is 18,000 deg$^2$, we however note that the best part of LSST's footprint for extragalactic studies might be smaller and correspondingly the required time WFIRST would spend on a wide band survey would be shorter. Such a wide survey component could be implemented as part of the nominal HLS survey, or as part of the HLS and other WFIRST survey components (depending on their interest); it could also be conducted as part of an extended mission, which would map nicely onto the LSST Y10 timescale.  

By the time WFIRST launches, LSST will have been in survey operation mode for several years already and will have built up substantial survey depth and area. If blending is an issue for LSST shape and photo-z measurements, WFIRST could provide a significant contribution to a possible solution in just 5 months. It is also possible that the lack of deep training data (spectra) for photo-z, or the lack of multi-band IR coverage will limit LSST photo-z accuracy. If the lack of spectroscopic information is limiting, exploring the idea of extended WFIRST grism observations across the LSST footprint might also be a good option. If narrow band IR information is useful, a WFIRST wide survey with the H-band is interesting to study further. A WFIRST H-band survey is approximately 3.5 times as slow as a W-band survey but it significantly limits wavelength dependent PSF problems. These ideas illustrate the flexibility of the WFIRST as an observatory, which will benefit multiple science cases across the cosmological community in the next decade.

Exploring optimal joint science strategies for WFIRST and LSST requires complex calculations and meaningful metrics. This paper has presented a connected infrastructure of survey simulations and sophisticated likelihood analyses that can be used to further explore joint LSST+WFIRST science cases. A WFIRST wide data set would impact multiple cosmological probes beyond 3x2pt cosmology and we mention cluster cosmology, voids, and trough cosmology, strong lensing, CMB lensing, and SZ synergies as examples for future consideration. 

Including these observables in a multi-probe analysis and improving the modeling of systematics are meaningful extensions of the work presented here. In particular, we plan to include catastrophic redshift outliers and higher order galaxy bias models that allow pushing to smaller scales in galaxy-galaxy lensing and galaxy clustering, in future analyses.  

\section*{Acknowledgments}
\copyright 2020. All rights reserved. This work is supported by NASA ROSES ATP 16-ATP16-0084 and NASA 15-WFIRST15-0008 grants. Support for MS was provided by the University of California Riverside Office of Research and Economic Development through the FIELDS NASA-MIRO program. The Flatiron Institute is supported by the Simons Foundation. Simulations in this paper use High Performance Computing (HPC) resources supported by the University of Arizona TRIF, UITS, and RDI and maintained by the UA Research Technologies department. Part of the research described in this paper was carried out at the Jet Propulsion Laboratory, California Institute of Technology, under a contract with the National Aeronautics and Space Administration. 

\bibliographystyle{mnras}
\bibliography{references.bib}
\label{lastpage}
\end{document}